\documentclass[12pt]{amsart}

\usepackage{amsaddr}
\usepackage{amssymb}
\usepackage{enumerate}
\usepackage{enumitem}
\usepackage{amsmath,amsthm,marvosym,wasysym,mathdots}
\usepackage{bbm}
\usepackage{natbib}
\usepackage{color}
\usepackage{setspace}
\usepackage{epsfig}
\usepackage{hyperref}
\usepackage{subcaption}
\usepackage{pbox}
\usepackage[stable]{footmisc}
\hypersetup{%
  colorlinks=false,% hyperlinks will be black,
  pdfborder = {0 0 0.5 [3 0]}
}
\newcommand{\indic}[1]{1\hspace{-2.1mm}{1}_{\{#1\}}} %Indicator Function
\usepackage{caption}
\usepackage{multirow}
\usepackage[left=1in,right=1in,top=1.1in,bottom=1in]{geometry}

\DeclareMathOperator*{\esssup}{ess\,sup}
\DeclareMathOperator*{\essinf}{ess\,inf}

\DeclareMathOperator*{\sign}{sign}

\def\L{{\mathcal L}}

\newtheorem{theorem}{Theorem}
\newtheorem*{theorem*}{Theorem}

\newtheorem{definition}[theorem]{Definition}

\newtheorem{example}[theorem]{Example}

\newtheorem{lemma}[theorem]{Lemma}

\newtheorem{proposition}[theorem]{Proposition}
\newtheorem{remark}[theorem]{Remark}

\theoremstyle{remark}
\newtheorem*{rem}{Remark}

\theoremstyle{definition} 
\newtheorem*{definition*}{Definition}

\def\N{{\mathbb N}} % positive integers
\def\Q{{\mathbb Q}} % rationals
\def\R{{\mathbb R}} % reals
\def\P{{\mathbb P}} % probability
\newcommand{\EE}{{\mathord{I\kern -.33em E}}}
\def\E{{\mathbb E}} % expectation
\def\1{1{\hskip -3.3 pt}\hbox{I}}
\def\F{{\mathcal F}}

\def\L{{\mathcal L} \,}

\providecommand{\varitem}{}


\numberwithin{equation}{section}
\numberwithin{theorem}{section}

\usepackage{float}
\usepackage{hhline}
\usepackage[font=small,skip=0pt]{caption}
\usepackage{bm}
\usepackage{pbox}
\usepackage{hyperref}
\numberwithin{equation}{section}
\numberwithin{theorem}{section}

\usepackage{makecell}
\usepackage{algorithm} 
\usepackage{algpseudocode} 
\algdef{SE}[DOWHILE]{Do}{doWhile}{\algorithmicdo}[1]{\algorithmicwhile\ #1}%
\algblock{Inputs}{EndInput}
\algnotext{EndInput}
\algblock{Output}{EndOutput}
\algnotext{EndOutput}

\makeatletter\def\@font@warning#1{}\makeatother

\title{Extreme Measures in
 Continuous Time Conic Finance}
\author{{Y}oshihiro {S}hirai} \date{\today}
\address{University of Maryland, College Park, Department of Mathematics}
\email{yshirai@umd.edu}


\begin{document}

%\begin{LARGE}

%\end{nouppercase}
%\maketitle

\begin{abstract}
Dynamic spectral risk measures define a claim's valuation bounds as supremum and infimum of expectations of the claim's payoff over a dominated set of measures. The measures at which such extrema are attained are called \textit{extreme} measures. We determine explicit expressions for their Radon-Nykodim derivatives with respect to the common dominating measure. Based on the formulas found, we estimate the extreme measures in two cases. First, the dominating measure is calibrated to mid prices of options and valuation bounds are given by options bid and ask prices. Second, the dominating measure is estimated from historical mid equity prices and valuation bounds are given by historical 5-day high and low prices. In both experiments, we find that the market determines upper bounds by testing scenarios in which losses are significantly lower than expected under the dominating measure, while lower bounds by ones in which gains are only slightly lower than in the base case. 
% In addition, we introduce continuous time Convex Finance by defining non homogeneous valuation bounds based on a convex set of acceptable risks that is not a cone.
\end{abstract}

\maketitle
%\tableofcontents

%\pagebreak

\section{Introduction}

Much of the existing literature on continuous time valuation bounds defines upper and lower prices as suprema and infima of conditional expectations of discounted payoffs over a certain set $\mathcal{M}$ of measures. When $\mathcal{M}$ is weakly compact and the bounds are time consistent, the extrema are attained at the same measure at different points in time (\cite{Delbaen-mstable06}). We call such measures \textit{extreme} for analogy with those analyzed in \cite{ChernyExtremeMeasures} in a static setting. The main mathematical contribution of this paper is to construct a pair of extreme measures for the continuous time Conic Finance bounds defined in \cite{MadanPistoriusStadje}.\footnote{The point of view in \cite{MadanPistoriusStadje} is that of risk measures, whereas here it is on valuation bounds. The mathematical aspects of the two theories are the same (see e.g. \cite{JashckeKuchler}).}

The fundamental assumption of Conic Finance, introduced in \cite{MadanCherny}, is that risks cannot be fully hedged and so the set of trades deemed \textit{acceptable} by market operators must strictly contain that of arbitrages. Acceptability is defined in Conic Finance by assuming that the market chooses a reference probability space $(\Omega,\F,\Q)$ and a set of test measures $\mathcal{M}$ dominated by $\Q$, so that a payoff is acceptable if its expected value under any test measure is nonnegative. Other definitions include those of \cite{Ledoit}, \cite{CochraneSaaRequejo}, \cite{ChernyHodges} and \cite{BernardoLedoit}. Furthermore, valuation bounds can also be defined based on indifference pricing (see \cite{CarmonaBook}) and model-free hedging (\cite{Hobson}).

Each of these alternatives presents its own advantages. As explained in \cite{MadanCherny}, Conic Finance valuations are independent from agents' preferences and initial wealth and they are robust to model misspecifications. In addition, generating upper and lower prices in Conic Finance does not require the existence of underlying liquid securities. 

From a mathematical perspective, upper and lower valuations are defined in continuous time Conic Finance as the unique solutions of respective upper and lower backward stochastic differential equations (BSDEs) driven by a Choquet integral and with pure jump Levy generator $X$ (see \cite{MadanPistoriusStadje}). In this case, and to the author's knowledge, a full proof for general formulas identifying the Radon-Nikodym derivatives of a pair of extreme measures $\overline{\Q}$ and $\underline{\Q}$ with respect to $\Q$ is absent from the literature. This paper's mathematical contribution is then Theorem \ref{SMMformula}, in which such an explicit expression is provided in terms of the level sets of the control processes $\overline{Z}$ and $\underline{Z}$ of the upper and lower BSDEs.

By the formulas obtained, the bounds defined by continuous time Conic Finance are no longer linear, as in static Conic Finance, over a pair of comonotonic payoffs. The requirement for linearity is now that the control processes associated to the value of the two claims be comonotonic for \textit{all} $t\in[0,T]$, where $T$ is the expiration date, and we show in Remark \ref{ComRem} that this is indeed more restrictive than just comonotonicity of payoffs. This is no surprise: by Theorem 7.1 in \cite{DelbaenCommTC}, a comonotonic and time consistent dynamic expectation on the set of bounded random variables must be a conditional expectation. Another interesting consequence of our formulas is that, differently from the static case (\cite{kusuoka}), continuous time Conic Finance valuations are not law invariant. More precisely, as observed in Remark \ref{LawInvariance}, equivalence of the valuation bounds of two claims requires the L\'evy measure of $X$ under $\Q$ of the $\alpha$-level sets of $\overline{Z}_t$ and $\underline{Z}_t$ to be the same for each level $\alpha$ and \textit{all} $t\in[0,T]$. Such result is in line with the one proved in  \cite{KupperSchachermeyer} that the only time consistent, law invariant, dynamic nonlinear expectation is the entropic risk measure.

%The proof of Theorem \ref{SMMformula} relies on the fact that the driver functions of the BSDEs are Choquet integrals and thus linear over comonotone integrands. As a consequence, this proof could be applied for $X$ a Sato or Hunt process, as well as to BSDEs with state and/or time dependent Choquet drivers (provided of course that a solution to such BSDE exists). The exposition in this paper is limited to a L\'evy setting as, in this case, the BSDEs' solutions exist and are unique (see \cite{BarlesBuckdahnPardoux}). 

In our L\'evy setting, the existence of the extreme measures implies that, as shown in \cite{BarlesBuckdahnPardoux}, there are functions $u,\ell:[0,T]\times \R^D\backslash\{0\}\rightarrow \R$ such that $u(t,X_t)$ and $\ell(t,X_t)$ are the upper and lower valuations for each $t\in[0,T]$. Furthermore, as shown in \cite{Denneberg}, the bounds are Malliavin differentiable and so the control processes satisfy
\begin{equation}\label{Malliavin}
\begin{aligned}
\overline{Z}_t(y)&=u(t,X_{t-}+y)-u(t,X_{t-}), \
\underline{Z}_t(y)=\ell(t,X_{t-}+y)-\ell(t,X_{t-})
\end{aligned}
\end{equation}
for every $t\in[0,T]$. Based on (\ref{Malliavin}), we show in Theorem \ref{MonotoneMeasures} that if $X$ has dimension $1$ and $f$ is monotone, the level sets of the control processes are time independent and deterministic, and so $X$ is a L\'evy process under $\overline{\Q}$ and $\underline{\Q}$. 

This result paves the way for two empirical studies that constitute the second and third contributions of this paper. In the first such study, the law of $X$ under the reference measure $\Q$ is obtained by calibrating the bilateral gamma (BG) process introduced in \cite{KuchlerTappe} to mid prices of options on the SPY exchange trade fund (SPY). Since options' payoffs are monotonic, $X$ is a L\'evy process under the extreme measures associated to options' valuation bounds, and the respective upper and lower L\'evy measures can be calibrated to bid and ask prices of options via FFT (see \cite{CarrMadanFFT}). Our goals are:
\begin{itemize}
\item[1.] to assess if continuous time Conic Finance bounds can match relative bid-ask spreads across strikes; and
\item[2.] to infer, based on the different relative bid-ask spreads across strikes, which events the market is more uncertain about.
\end{itemize} 
The importance of 1. is that the calibration exercise determines the set $\mathcal{M}$ of test measures. If the relative options bid-ask spreads are matched, the same set $\mathcal{M}$ can be used to generate quotes for certain derivatives traded out of the counter, such as straddles and, more in general, combo options. This is an important issue for market makers as they need to quickly provide quotes that are cheap but accurate enough. With this consideration in mind, we also find conditions on the set $\mathcal{M}$ for $X$ to be a BG process under both extreme measures and assuming it is a BG process under $\Q$. The conditions obtained resemble the Wang transform (\cite{Wang2000}, \cite{Wang2002b}), with the BG distribution replacing the Gaussian. When they are enforced, calibration is further facilitated, although the error is higher (see Remark \ref{CalibBG2BG}).

The importance of 2. is that the ability to explain why some events are more uncertain than others based on market data is one way to assess the calibrated set $\mathcal{M}$. We set $X=G-L$, where $G$ and $L$ are positive gamma processes referred to as gains and losses (see \cite{KuchlerTappe} for their existence). Then, we find that bounds are determined by distorting the loss process $L$ for the ask price of calls and the bid price of puts, and the process $G$ for the bid price of calls and ask price of puts. That is, a call's ask and a put's bid are determined by testing their payoff under scenarios in which there is a high chance that $L_T$ is lower than expected under $\Q$. For a put's ask or a call's bid, instead, the payoff is tested against the event that $G_T$ may be higher than expected. We then find that the stress on the loss process is higher than that on the gain process. This is related to the investors' need to hedge against economic downturns, which makes the out of the money (OTM) puts market more liquid than that of OTM calls. Only few existing empirical studies are performed on options bid and ask prices, so this contribution is quite unique in the empirical finance literature.

For the second empirical study of this paper we assume that, under the reference measure, the law of $X$ is again BG but this time estimated from daily closing mid prices of the SPY. Furthermore, the observed upper and lower valuations are defined by the SPY 5-day high and low prices. This is based on the fact that large trades put in place by institutional investors are often executed over a few days at least. The payoff of owning the SPY is defined by $Y_0e^{X_T}$, where $Y_0$ is the current value of the SPY and $T$ is 5 days. Then $X$ is again a L\'evy process under $\overline{\Q}$ and $\underline{\Q}$ and its L\'evy measure can be expressed in terms of an integral. Hence, the probability density of $X_T$ can be calculated by Fourier inversion, and estimated by matching its tail to that of the empirical distribution (as in \cite{MadanEstimate}). The resulting estimator is called the digital moment estimator (DM). Our goals are:
\begin{itemize}
[nolistsep,noitemsep]
\item[1.] to compare the estimates we obtain from DM with the one obtained through the generalized method of moment (GMM); and
\item[2.] to infer, based on historical equity prices, which events market operators are more uncertain about.
\end{itemize}
The importance of 1. is to determine the usefulness of our formulas for the Radon-Nykodim derivatives for estimation purposes. In fact, computation of the tail measure is a demanding task without knowledge of the probability density of $X_T$, prone to numerical error and approximations. Hence, without the formulas developed in this paper, one is forced to use methods, such as the GMM, that are not designed to incorporate in their estimates events that occur less frequently. The importance of 2. is as in the study on options. We find that:
\begin{itemize}
[nolistsep,noitemsep]
\item[1.] with DM estimators, and as found in our first study on options data, upper valuations are determined by uncertainty in the loss process and lower valuations by uncertainty in the gain process; GMM estimators are, instead, more balanced;
\item[2.] the drifts of the lower valuation process estimated by the two methods are similar; however, the DM drift of the upper valuation is lower than the GMM one;
\item[4.] the correlation between the DM estimated upper drivers and upper returns is similar to that between lower driver and lower returns; for GMM, it is substantially higher;
\item[3.] as in the case of calibration to options, the higher DM estimated upper driver implies that the market tests scenarios in which losses are much lower than expected, and gains only a bit higher; this appears related to the monetary authority's support to the financial sector and the real economy during the 2010-2020 decade;
\item[5.] the standard deviation of the upper valuation implied by DM estimators is higher, on average, than that of the density of the lower valuation; they are similar for GMM estimated densities.
\end{itemize}

From these observations we argue that market operators were more uncertain over the period considered about the SPY's loss process than its gain process. Furthermore, the GMM fails to incorporate in its estimates extreme realizations of upper and lower returns.
The rest of the paper is organized as follows. In section 2 we review Conic Finance valuation bounds and prove preliminary results. The formula for the Radon-Nikodym derivatives $d\overline{\Q}/d\Q$ and $d\underline{\Q}/d\Q$ is given in section 3. Section 4 considers the case of monotonic claims. Results on numerical experiments are reported in Section 5 and 6. Section 7 concludes.

\section{Assumptions and Preliminary Results}\label{setup}

\subsection{Assumptions} Unless otherwise specified, the following assumptions, most of which are as in \cite{MadanPistoriusStadje}, hold throughout the rest of the paper. 
\begin{itemize}
[nolistsep,noitemsep]
\item[(i)] Given a topological space  $(\mathcal{X},\tau)$, $\mathcal{B}(\mathcal{X})$ denotes the Borel $\sigma$-algebra on $\mathcal{X}$.
\item[(ii)] Given a measurable space $(\Omega,\F)$, random processes $X^i=\{X^i_t\}_{t\geq 0}$ on $(\Omega,\F)$ and constants $Y_0^i$, $i=1,...,D$, and $T>0$, we consider a market composed of a risk-free asset with rate $r\geq 0$ and $D$ risky assets with payoff $Y^i_T:=Y^i_0e^{X^i_T}$. To simplify notation, the rate $r$ is normalized to zero, except in the empirical studies (Sections 5 and 6). Furthermore, we set $X:=(X_1,...,X_n)$ and $\{\F_t\}_{t\geq 0}$ denotes the right continuous, completed filtration generated by $X$.
\item[(iii)] There is a probability measure $\Q$ on $(\Omega,\F)$ such that, for each $i$, $Y^i_T\in L^2(\Omega,\F,\Q)$ and the discounted process $Y^i=\{Y^i_t\}_{t\geq 0}$ defined by $Y^i_t:=Y_0^ie^{-rt+X_t}$ is a martingale under $\Q$. This is a condition necessary to avoid arbitrages (\cite{JouiniKallal}, Theorem 2.1).
\item[(iv)] The process $X$ satisfies, for $t\geq 0$ and $\mathtt{d}\in\R^{D}\setminus\{0\}$,
\begin{align*}
X_t=\mathtt{d} t+\int_{[0,t]\times\R^D\backslash\{0\}}y \tilde{N}(dy,ds).
\end{align*}
where $\tilde{N}(dy,ds):=N(dy,ds)-\nu(dy)ds$ and $N$ is a Poisson random measure with $\Q$-compensator $\nu$. It is assumed that $\nu$ is $\sigma$-finite, has no atoms and, for some $\varepsilon>0$,
\begin{align*}
\int_{\R^D\backslash\{0\}}|y|^{2+\varepsilon}\nu(dy)\in\R^D\backslash\{0\}.
\end{align*}
\item[(v)] $\{X^{t,x}_s\}_{s\geq t}$ is defined for $s\geq t\geq 0$ and $x\in\R^D$ by
\begin{align*}
X^{t,x}_s=x+\mathtt{d}(s-t)+\int_{[t,s]\times \R^D\backslash\{0\}}y \tilde{N}(dy,ds).
\end{align*}
\item[(vi)] $\Gamma=(\Gamma_+,\Gamma_-)$ is a pair of measure distortions, i.e. $\Gamma_+,\Gamma_-:[0,\nu(\R))\rightarrow \R_+$ are bounded, concave and satisfy $\Gamma_-(x)\leq x$ and
\begin{align}\label{MDCondition}
\int_{(0,\nu(\R))}\frac{\Gamma_{\pm}(y)}{2y^{3/2}}dy < \infty.
\end{align}
\item[(vii)] $L^2(\nu):=L^2(\R\backslash\{0\},\mathcal{B}(\R\backslash\{0\}),\nu)$ and $g:L^2(\nu)\rightarrow \R$ is specified for $z\in L^2(\nu)$ by
\begin{align}\label{Driver}
g(z):=\int_0^{\infty}\Gamma_+(\nu(z^+>a))da+\int_0^{\infty}\Gamma_-(\nu(z^->a))da.
\end{align}
\end{itemize}
%The measure distortions $\Gamma^+,\Gamma^-$ define the driver function $g$ in continuous time Conic Finance in a similar way as probability distortions define a driver for the BS$\Delta$E satisfied by the valuation bounds of discrete time Conic Finance valuation bounds (see \cite{EMPSY}). Our main results are stated in terms of any pair of measure distortions $\Gamma_+$ and $\Gamma-$ that satisfy the above assumptions, but in applications they will often be specified as in Example \ref{MeasureDistortionEx}.

\begin{example}\label{MeasureDistortionEx} An example of measure distortions is the pair $\Lambda=(\Lambda_+,\Lambda_-)$ defined by
\begin{equation}\label{MeasureDistortion}
\begin{aligned}
\Lambda_+(x)
	  &:=a\left(1-e^{-cx}\right)^{1/(1+\gamma)},  \
\Lambda_-(x)
	   := \frac{b}{c}\left(1-e^{-cx}\right),
\end{aligned}
\end{equation}
where $x\geq 0$, $0<\gamma<1$, $0< b\leq 1$ and $a,c>0$. These distortions are obtained in \cite{EMPY} by composing common probability distortions with the change of variable $x\rightarrow 1-e^{-cx}$. Note that if $\gamma\geq 1$ then assumption \ref{MDCondition} does not hold, and if $b>1$ there is $x>0$ such that $\Gamma_-(x)>x$. The requirement $\gamma>0$ ensures that the associated probability distortion is strictly concave. See \cite{EMPSY} for a description of the parameters $a,b,c,\gamma$.
%If an event $A$  satisfies $\nu(A)\geq \tfrac{10}{c}$, then $\exp(-c(\nu(A)) \approx 0$, and, essentially, $\nu(A)$ is not reweighted.
%The parameter $\gamma$ determines the convexity of $\Gamma_+$ around the origin, so the higher is $\gamma$, the higher the reweighting for large gains (for the upper valuation) or losses (for the lower one), and the higher the distortive effects of $\Gamma_+$.
\end{example}

\subsection{Notation}\label{Notation} In addition to the assumptions above, the following notation will be used throughout the paper.

\begin{itemize}
\item[i.] For $p\in [1, \infty]$, $L^p:=L^p(\Omega,\F_T,\Q)$; recall that $L^1\supset L^2\supset... \supset L^{\infty}$.
\item[ii.] $L^2(\nu)$ is endowed with the Borel $\sigma$-algebra generated by the $L^2(\nu)$-norm topology.
\item[iii.] For an $L^2(\nu)$-valued process $\{V_t\}_{t\geq 0}$ and $y\in\R^D\backslash\{0\}$, we often write $V_t(y)$ for $V_t(\omega,y)$.
\item[iv.] $\mathcal{E}$ denotes the Doleans-Dade exponential.
\item[v.] $\mathcal{P}$ denotes the predictable $\sigma$-algebra on $[0,T]\times\Omega$.
\item[vi.] We often identify the set of test measures $\mathcal{M}$ with the subset of $L^1$ of their Radon Nikodim derivatives. A weak* topology on $\mathcal{M}$ can be defined as in Corollary 14.11 of \cite{Aliprantis}. This topology is equivalent to the weak topology on $L^1$, i.e. the topology such that if $\{\chi_n\}_{n\in\N}\subset \mathcal{M}$, then $\chi_n\rightarrow \chi\in \mathcal{M}$ if and only if 
\begin{align}\label{weak1}
\int_{\Omega} C(\omega)\chi_n(d\omega)
\rightarrow \int_{\Omega} C(\omega)\chi(d\omega), \ \forall C\in L^{\infty}.
\end{align}
If $\mathcal{M}\subset L^2$, we also have the weak topology in $L^2$, i.e. (\ref{weak1}) must hold for all $C\in L^2$. Recall also that, by the Eberlein-Smulian theorem, weak compactness in $L^p$ is equivalent to weak sequential compactness in $L^p$, $1\leq p\leq \infty$.
\item[vii.] $C(\Gamma)$ denotes the set of functions $q\in L^2(\nu)$ s.t., for $A\in \mathcal{B}(\R^D\backslash\{0\})$ with $\nu(A)<\infty$,
\begin{align}\label{psixi}
-\Gamma_-(\nu(A))\leq \int_Aq(y)\nu(dy)\leq \Gamma_+(\nu(A)).
\end{align}
\item[viii] For $\Gamma_+,\Gamma_-$ differentiable  and $D=1$, we set
\begin{equation}\label{psifun}
\begin{aligned}
\overline{\psi}_{\Gamma}(y)
	& := \Gamma_+'\left(\nu([y,\infty))\right)\indic{y>0}
    		-\Gamma_-'\left(\nu((-\infty,y]))\right)\indic{y<0}, \\
\underline{\psi}_{\Gamma}(y)
& :=  -\Gamma_-'\left(\nu([y,\infty))\right)\indic{y>0}
	+\Gamma_+'\left(\nu((-\infty,y]))\right)\indic{y<0}.
\end{aligned}
\end{equation}
\item[ix] For $\Gamma_+,\Gamma_-$ differentiable and $D=1$, $\overline{\Q}(\Gamma)$ and $\underline{\Q}(\Gamma)$ denote test measures under which the compensator of $N$ is respectively given by
\begin{align*}
(1+\overline{\psi}_{\Gamma}(y))\nu(dy), \
(1+\underline{\psi}_{\Gamma}(y))\nu(dy)
\end{align*}
\end{itemize}
Furthermore, given a function $u:[0,T]\times \R^D\rightarrow \R$ and $z:\R^D\backslash\{0\}\rightarrow \R$,
\begin{itemize}
\item[x.] $\mathcal{D}_u^{t,x}(y):=u(t,x+y)-u(t,x)$ for all $(t,x,y)\in[0,T]\times\R^D\times\R^D\setminus\{0\}$;
\item[xi.] $A^\alpha_{z}=\{y\in\R^D\backslash\{0\}:z(y)\geq \alpha\}$ if $\alpha>0$;
%\begin{align*}
%A^{\alpha}_{t,x}=\{y\in\R^D\backslash\{0\}:z(t,x,y)>\alpha\};
%\end{align*}
\item[xii.] $A^\alpha_{z}=\{y\in\R^D\backslash\{0\}:z(y)\leq \alpha\}$ if $\alpha<0$;
%\begin{align*}
%A^{\alpha}_{t,x}=\{y\in\R^D\backslash\{0\}:z(t,x,y)<\alpha\};
%\end{align*}
\item[xiii.] $z^{\alpha}(y):=\sign(\alpha)\indic{A^{\alpha}_{z}}(y)$ for $\alpha\in\R^D\backslash\{0\}$ and $y\in\R^D\backslash\{0\}$;
\item[xiv.] $\Sigma_{z}$ denotes the completed $\sigma$-algebra generated by $z$;
\item[xv.] $\nu_{z}$ denotes the restriction of $\nu$ to $\Sigma_{z}$.
\end{itemize}

\subsection{Valuation Bounds and Associated BSDEs}
\begin{definition}\label{NonlinearVal}
For each $\mathcal{P}$-measurable, $C(\Gamma)$-valued process $\psi:=\{\psi_s\}_{t\in [0,T]}$, let $M^{\psi}:=\{M^{\psi}_t\}_{t\in[0,T]}$ be defined by
\begin{align*}
M^{\psi}_t:=\int_{[0,t]\times \R^D\backslash\{0\}}\psi_s(y)\tilde{N}(ds,dy)
\end{align*}
Let $\mathcal{M}$ denote the set of all measures absolutely continuous with respect to $\Q$ and with square integrable Radon-Nikodym derivative $\xi$ such that $\xi=\mathcal{E}(M^{\psi})_T$ for some $\mathcal{P}$-measurable, $C(\Gamma)$-valued process $\psi:=\{\psi_s\}_{t\in [0,T]}$. The upper and lower valuations of a claim $C\in L^2$ to be delivered at time $T$ are the processes $U=\{U_t\}_{t\geq 0}$ and $L=\{L_t\}_{t\geq 0}$ defined, for $t\in[0,T]$, by
\begin{align}\label{UandL}
U_t:=\esssup_{\Q^{\psi}\in \mathcal{M}}\E^{\Q^{\psi}}[C|\F_t], \
L_t:=\essinf_{\Q^{\psi}\in \mathcal{M}}\E^{\Q^{\psi}}[C|\F_t].
\end{align}
\end{definition}
The processes $U$ and $L$ are solutions of a BSDE with driver given by the functional $g$ defined by (\ref{Driver}). This result is shown in \cite{LaevenStadje} for a general class of nonlinear utility functions. The key connection with our setting is provided by the following characterizations of the driver function $g$ and of its subdifferential.
\begin{proposition}\label{driver_prop} The functional $g$ defined by (\ref{Driver}) satisfies
\begin{align}\label{driver}
g(z)
=\sup_{q\in C(\Gamma)}
	\int_{\R^D\backslash\{0\}}q(x)z(x)\nu(dx),
\end{align}
for every $z\in L^2(\nu)$.
\end{proposition}
\begin{proof}
This result is well known when $\nu$ is finite, and its extension to the non finite case is obtained by approximating $\nu$ by finite measures. See Proposition 3.5 in \cite{MadanPistoriusStadje}.
\end{proof}
For the next result, recall that the subgradient $\partial g(z)$ of $g$ at $z\in L^2(\nu)$ is defined by
\begin{align}\label{subdiff}
\partial g(z):=
\left\lbrace q \in L^2(\nu): 
	\int_{\R^D\backslash\{0\}}q(y)
		\left(\tilde{z}(y)-z(y)\right)\nu(dy)
	\leq g(\tilde{z})-g(z) \ \forall \tilde{z}\in L^2(\nu)\right\rbrace.
\end{align}
\begin{proposition}\label{MartingaleMeasure}
Let $z\in L^2(\nu)$. Then, there is $q\in C(\Gamma)$ such that 
\begin{align}\label{subgr1}
g(z) = \int_{\R^D\backslash\{0\}} q(y)z(y)\nu(dy)
\end{align}
and, for every $\tilde{z}\in L^p(\nu)$,
\begin{align}\label{subgr2}
\int_{\R^D\backslash\{0\}} q(y)\tilde{z}(y)\nu(dy) \leq g(\tilde{z}).
\end{align}
In addition, the set of functions $q\in C(\Gamma)$ satisfying (\ref{subgr1}) and (\ref{subgr2}) coincides with $\partial g(z)$.
\end{proposition}
\begin{proof}
Define the 1-dimensional vector space $\Theta:=\{az\}_{a\in\R}$
and a functional $\iota:\Theta\rightarrow \R$ by $\iota[az]:=ag(z)$. Since $\iota$ is linear on $\Theta$ and dominated by $g$, the existence of $q$ follows by the Hahn-Banach and Riesz representation theorems. As for the last statement, it is clear that if $q\in C(\Gamma)$ satifies (\ref{subgr1}) and (\ref{subgr2}) then $q\in \partial g(z)$. On the other hand, if $q \in \partial g(z)$,
\begin{align}\label{leq}
\int_{\R^D\backslash\{0\}}q(y)z(y)\nu(dy) \geq g(z),
\end{align}
which follows from setting $\tilde{z}=0$ in (\ref{subdiff}). Hence,
\begin{align*}
\int_{\R^D\backslash\{0\}}q(y)\tilde{z}(y)\nu(dy) \leq g(\tilde{z})
\end{align*}
for every $\tilde{z}\in L^2(\nu)$, and, setting $\tilde{z}=\indic{A}$ for $A\in \mathcal{B}(\R^D\backslash\{0\})$, we obtain (\ref{psixi}), and so $q \in C(\Gamma)$. But then (\ref{driver}) implies
\begin{align*}
g(z)\geq \int_{\R^D\backslash\{0\}}q(y)z(y)\nu(dy),
\end{align*}
which, combined with (\ref{leq}) yields the result.
\end{proof}

\begin{theorem}\label{BSDETh} Let $C=f(X_T)\in L^2$. Then, there are $L^2(\nu)$-valued predictable processes $\overline{Z}=\{\overline{Z}_t\}_{t\in[0,T]}$ and $\underline{Z}=\{\underline{Z}_t\}_{t\in[0,T]}$ such that the quantities
\begin{align*}
\E^{\Q}\left[\sup_{t\in[0,T]}\left(\int_{\R^D\backslash\{0\}} \overline{Z}^2_t(y)\nu(dy)\right)^{1/2}\right], \
\E^{\Q}\left[\sup_{t\in[0,T]}\left(\int_{\R^D\backslash\{0\}} \underline{Z}^2_t(y)\nu(dy)\right)^{1/2}\right]
\end{align*}
are finite and $(U,\overline{Z})$ and $(L,\underline{Z})$, with $U$ and $L$ defined by (\ref{UandL}), are the unique solutions of
\begin{align}
U_t
&=C+\int_t^Tg(\overline{Z}_s)ds
	-\int_{(0,T]\times \R^D\backslash\{0\}}\overline{Z}_s(y)\tilde{N}(ds,dy), \label{BSDE}\\
L_t
&=C-\int_t^Tg(\underline{Z}_s)ds
	-\int_{(0,T]\times \R^D\backslash\{0\}}\underline{Z}_s(y)\tilde{N}(ds,dy). \label{BSDEL}
\end{align}
Furthermore, there are measurable functions $u,\ell:[0,T]\times\R^D\rightarrow \R$ such that $u(t,X_t)=U_t$ and $\ell(t,X_t)=L_t$ for every $t\in [0,T]$.
\end{theorem}
\begin{proof}
The proof is based on showing that the predictable component in the Doob-Meyer decompositions of $U$ and $L$ is the integrated driver function. See \cite{LaevenStadje}, Theorem A.25. The last statement follows from Corollary 2.3 in \cite{BarlesBuckdahnPardoux}.
\end{proof}
From the proof of Theorem \ref{BSDETh} we also obtain a few important properties of the set $\mathcal{M}$ and the valuation bounds $U$ and $L$ of a claim $C$. 
\begin{theorem}\label{remSMM}
Let $\mathcal{M}$ be as in Definition \ref{NonlinearVal} and $C$, $\overline{Z}$, $\underline{Z}$, $U$, $L$ as in Theorem \ref{BSDE}. Then:
\begin{itemize}
\item[i.] The set $\mathcal{M}$ is weakly compact in $L^2$;
\item[ii.] There is a pair of predictable selectors $(\overline{\psi},\underline{\psi})=\{(\overline{\psi}_t,\underline{\psi}_t)\}_{t\geq 0}$ of $\partial g(\overline{Z}_t)$ and $\partial g(\underline{Z}_t)$ for $t\in [0,T]$, i.e. 
\begin{align*}
\overline{\psi}_t\in \partial g(\overline{Z}_t), \
\underline{\psi}_t\in \partial g(\underline{Z}_t), \
t\in[0,T];
\end{align*}
\item[iii.] The measures $\Q^{\overline{\psi}},\Q^{\underline{\psi}}\in \mathcal{M}$ associated to any pair $(\overline{\psi},\underline{\psi})$ in ii. are well defined and 
\begin{align*}
U_t=\E^{\Q^{\overline{\psi}}}[C|\F_t], \
L_t=\E^{\Q^{\underline{\psi}}}[C|\F_t], \
t\in[0,T].
\end{align*}
\end{itemize}
\end{theorem}
\begin{proof}
The proof of ii., as mentioned in the proof of Theorem A.25 in \cite{LaevenStadje}, follows by applying the Kuratowski and Ryll-Nardzewski measurable selection theorem, while iii. is Lemma A.24 in \cite{LaevenStadje}. The proof of i. is obtained by first identifying $\mathcal{M}$ with the collection $\{d\Q'/d\Q\}_{\Q'\in\mathcal{M}} \subset L^2$ of the Radon-Nykodim derivatives of its components. Now, for every $C\in L^2$, since $C$ is $\F_T$-measurable, there is a Borel measurable function $f$ such that $C=f(X_T)$. By iii., there are predictable selectors $(\overline{\psi},\underline{\psi})$ and measures $\Q^{\overline{\psi}}$ and $\Q^{\underline{\psi}}$ such that 
\begin{align*}
\E^{\Q^{\overline{\psi}}}[C]=\sup_{Q\in\mathcal{M}}\E^{\Q}[C], \
\E^{\Q^{\underline{\psi}}}[C]=\inf_{Q\in\mathcal{M}}\E^{\Q}[C].
\end{align*}
Weak compactness in $L^2$ then follows by James' theorem and the fact that $\mathcal{M}$ is convex and weakly closed in $L^1$ (\cite{FollmerSchied}), and thus also weakly closed in $L^2$.
\end{proof}
From Theorem \ref{remSMM}, the following Definition is well posed.
\begin{definition} Let $C\in L^2$. Any pair of measures $\Q^{\overline{\psi}}$ and $\Q^{\underline{\psi}}$ as in Theorem \ref{remSMM} are called, respectively, upper and lower extreme measures for $C$.
\end{definition}
\begin{remark}\label{LebCont} Since weak compactness in $L^2$ implies weak compactness in $L^1$, by the Dunford Pettis theorem, the valuation bounds defined by \ref{UandL} are Lebsegue continuous, i.e. they are continuous under convergence in probability for uniformly bounded sequences of claims. Then, by Theorem 7.1 in \cite{DelbaenCommTC}, they are linear over all bounded comonotone claims if and only if they are linear over all bounded claims. That is, if and only if there is a fixed measure $\Q^*$ that is an upper and lower extreme measure for each bounded claim $C$. We will prove in Section \ref{Monotone} that this is not the case (specifically, see Remark \ref{ComRem}).
\end{remark}
Our goal in the next section will be to determine an explicit formula for the measurable selectors in Theorem \ref{remSMM}. This allows one to identify the compensators of $\tilde{N}$ under the upper and lower extreme measures in terms of the processes $\overline{Z}$ and $\underline{Z}$. The characterization of such processes is then useful for operative purposes, and is provided next.

\begin{theorem}\label{ThBSDEPIDE} Let $f:\R^D\rightarrow \R$ be Lipschitz continuous, and consider the PIDE
\begin{align}\label{PIDE}
\begin{cases}
u_t+\mathcal{G}u+g(\mathcal{D}_u^{t,x})=0\\
u(T,x)=f(x)
\end{cases}
\end{align}
where $\mathcal{D}_u^{t,x}(y):=u(t,x+y)-u(t,x)$ and
\begin{align*}
\mathcal{G}(u)(t,x)= \mathtt{d}^T\nabla u(t,x)+\int_{\R^D\backslash\{0\}}\left(\mathcal{D}_u^{t,x}(y)-\nabla u(t,x)^Ty\right)\nu(dy).
\end{align*}
Let $C=f(X_T)$. Then, the function $u:[0,T]\times\R^D\rightarrow \R$ defined in Theorem \ref{BSDETh} is the unique viscosity solution of (\ref{PIDE}) among the class of solutions satisfying for every $t\in[0,T]$, $c>0$,
\begin{align*}
\lim_{|x|\rightarrow\infty}|u(t,x)|e^{-c\log^2(|x|)}=0.
\end{align*}
Furthermore, if $u\in C^{0,1}([0,T]\times\R)$ or the probability law of $X^{t,x}_s$ is absolutely continuous with respect to the Lebesgue measure, then the process $\overline{Z}$ in Theorem \ref{BSDETh} satisfies, for $(t,y)\in[0,T]\times\R^D\backslash\{0\}$,
\begin{align*}
\overline{Z}_t(y)=\mathcal{D}_u^{t,X_{t-}}(y).
\end{align*} 
An analogous result holds for the function $\ell$ in Theorem \ref{BSDETh}.
\end{theorem}
\begin{proof} By Theorem \ref{remSMM}, the solution $(U,\overline{Z})$ to (\ref{BSDE}) also solves the linear equation
\begin{align}
U_t
=C+\int_t^T\int_{\R^D\backslash\{0\}}\overline{\psi}_s(y)Z_s(y)\nu(dy)ds
	-\int_{(0,T]\times \R^D\backslash\{0\}}Z_s(y)\tilde{N}(ds,dy),
\end{align}
where $\{\overline{\psi}_t\}_{t\geq 0}$ is the measurable selector process in Theorem (\ref{remSMM}). The result then follows by Theorem 4.1.4 and Theorem 4.2.2 in \cite{Delong}.
\end{proof}

\section{Determination of the Measurable Selectors}

We recall the definition of comonotonicity, needed in the proof of Theorem \ref{SMMformula} below.
\begin{definition}\label{COMon}
Let $\Theta$ be any set. Two functions $f,g:\Theta \rightarrow \R$ are called comonotone if there are no pairs $\theta_1,\theta_2\in \Theta$ such that $f(\theta_1)<f(\theta_2)$ and $g(\theta_1)>g(\theta_2)$.
\end{definition}
Definition \ref{COMon} is based on proposition 4.5 in \cite{Denneberg}. Recall also that Choquet integrals are additive over comonotone functions (see \cite{Denneberg} proposition 5.1). In particular, by definition \ref{COMon}, if $f$ and $h$ are comonotone and $g$ and $h$ are comonotone, than $f+g$ and $h$ are comonotone, so Choquet integrals are additive over finite sets of pairwise comonotone functions.

\vspace{3mm}

We also recall the following result, whose proof is based on assumption \ref{MDCondition}.
\begin{lemma}\label{convex} The set $C(\Gamma)$ is convex and closed and bounded in $L^2(\nu)$, so that it is weakly compact in $L^2(\nu)$. Furthermore, the driver function $g$ is Lipschitz-continuous for the $L^2(\nu)$-norm.
\end{lemma}
\begin{proof} See \cite{MadanPistoriusStadje} and the references therein.
\end{proof} 

\begin{theorem}\label{SMMformula}
Fix a claim $C\in L^2$ and suppose that $\Gamma_+,\Gamma_-$ are differentiable on $(0,\infty)$. Let $(U,\overline{Z})$ and $(L,\underline{Z})$ be the solutions of (\ref{BSDE}) and (\ref{BSDEL}) respectively. Define measures $\overline{\Q}$ and $\underline{\Q}$ by setting, for $t\geq 0$,
\begin{align*}
\E\left[\left.\frac{d\overline{\Q}}{d\Q}\right\rvert\F_t\right] =\mathcal{E}(\overline{M})_t, \
\E\left[\left.\frac{d\underline{\Q}}{d\Q}\right\rvert\F_t\right] =\mathcal{E}(\underline{M})_t,
\end{align*}
where 
\begin{align*}
\overline{M}_t &:= \int_{[0,t]\times \R^D\backslash\{0\}}\overline{\psi}_s(y)\tilde{N}(ds,dy),\
\underline{M}_t  := \int_{[0,t]\times \R^D\backslash\{0\}}\underline{\psi}_s(y)\tilde{N}(ds,dy)
\end{align*}
and where $\overline{\psi}=\{\overline{\psi_t}\}_{t\geq 0}$ and $\underline{\psi}=\{\underline{\psi_t}\}_{t\geq 0}$ are defined by
\begin{equation}\label{psiformula}
\begin{aligned}
\overline{\psi}_t(y)
& := \Gamma_+'(\nu(\{\overline{Z}_t>\overline{Z}_t(y)\})\indic{\overline{Z}_t(y)>0}
	 -\Gamma_-'(\nu(\{\overline{Z}_t<\overline{Z}_t(y)\})\indic{\overline{Z}_t(y)<0},\\
\underline{\psi}_t(y)
& := -\Gamma_-'(\nu(\{\underline{Z}_t>\underline{Z}_t(y)\})\indic{\underline{Z}_t(y)>0}
	 +\Gamma_+'(\nu(\{\underline{Z}_t<\underline{Z}_t(y)\})\indic{\underline{Z}_t(y)<0},
\end{aligned}
\end{equation}
with $\Gamma_+':=\tfrac{d}{dx}\Gamma_+$, $\Gamma_-':=\tfrac{d}{dx}\Gamma_-$. Then, $\overline{\psi}$ and $\underline{\psi}$ are measurable selectors of $\partial g(\overline{Z}_t)$ and $\partial g(\underline{Z}_t)$ respectively for $t\in [0,T]$, and $\overline{\Q}$ and $\underline{\Q}$ are well defined upper and lower extreme measures for $C$. 
\end{theorem}

\begin{rem} For the notation $\nu_z$, $\Sigma_z$, $A^{\alpha}_z$ and $z_{\alpha}$ in the proof of Theorem \ref{SMMformula}, see Section \ref{Notation}.
\end{rem}

\begin{proof} We show in the first two steps below that $\overline{\psi}$ is a predictable selector of $\partial g(\overline{Z}_t)$ for $t\in [0,T]$. The respective result for $\underline{\psi}$ can be shown analogously. The proof will then follow, as shown in Step 3, from an application of Lemma A.24 in \cite{LaevenStadje}.

\vspace{3mm}

\textit{\underline{Step 1}. Fix $(t,\omega)\in [0,T]\times \Omega$, and let $z:=\overline{Z}_t(\omega)$. There is a $\nu_{z}$-a.e. unique function 
\begin{align*}
\psi_t(\omega,\cdot):\R^D\backslash\{0\}\rightarrow \R
\end{align*}
such that $\psi_t(\omega,\cdot)$ is $\Sigma_{z}$ measurable and, for every $\alpha>0$,
\begin{equation}\label{ConsTBP}
\begin{aligned}
\int_{A^{\alpha}_{z}}\psi_t(\omega,y)\nu(dy)&=g(z), \\
\int_{A^{-\alpha}_z}\psi_t(\omega,y)\nu(dy)&=-g(z_{-\alpha}(t,x,\cdot)).
\end{aligned}
\end{equation}
Furthermore, $\psi_t(\omega,\cdot)$ satisfies 
\begin{align}
g(z)&=\int_{\R^D\backslash\{0\}}\psi_t(\omega,y)z(y)\nu(dy). \label{TBP}
\end{align}
}

\noindent Proof of step 1.

\vspace{1mm}

From Definition \ref{COMon}, if $I\subset\R\backslash\{0\}$ is finite, the functions $\left\lbrace z,\{z_{\alpha}\}_{\alpha\in I}\right\rbrace$ are pairwise comonotone. Hence, for every $(a_0,\{a_{\alpha}\}_{\alpha\in I})\in \R^{1+|I|}$,
\begin{align*}
g\left(a_0z+\sum_{\alpha \in I}a_{\alpha}z_{\alpha}\right) 
= a_0g(z)+a_{\alpha}\sum_{\alpha\in I}g(z_{\alpha}).
\end{align*}
Let 
\begin{align*}
\Theta_I:=span\left\lbrace z,\{z_{\alpha}\}_{\alpha\in I}\right\rbrace,
\end{align*}
and consider the functional $\iota:\Theta_I\rightarrow \R$ defined, for $(a_0,\{a_{\alpha}\}_{\alpha\in I})\in \R^{1+|I|}$, by
\begin{align*}
\iota\left[a_0z+\sum_{\alpha \in I}a_{\alpha}z_{\alpha}\right]
=g\left(a_0z+\sum_{\alpha \in I}a_{\alpha}z_{\alpha}\right).
\end{align*}
Since $\iota$ is linear and dominated by $g$, the Hahn Banach theorem implies that there is $q_I\in L^2(\nu)$ such that (\ref{ConsTBP}) and (\ref{TBP}) hold and, for every $z\in\L^2(\nu)$,
\begin{align*}
g(z') 
&\geq \int_{\R^D\backslash\{0\}}q_{I}(y)z'\nu(dy).
\end{align*}
In particular, for every $A\in\mathcal{B}(\R^D\backslash\{0\})$ such that $\nu(A)<\infty$,
\begin{align*}
-\Gamma_-(\nu(A))
& = -\int_0^{\infty}\Gamma_-(\nu(\indic{A}>s)ds 
  = -g(-\indic{A})\\
& = \inf_{q\in C(\Gamma)}\int_{A}q(y)\nu(dy)
  \leq \int_Aq_I(y)\nu(dy)
  \leq \sup_{q\in C(\Gamma)}\int_{A}q(y)\nu(dy)\\
& = g(\indic{A})
  = \int_0^{\infty}\Gamma_+(\nu(\indic{A}>s)ds 
  = \Gamma_+(\nu(A)),
\end{align*}
so that $q_I\in C(\Gamma)$. 

Then, the sets $\{\Psi_{\alpha}\}_{\alpha\in\R\backslash\{0\}}$ defined, for $\alpha\in\R\backslash\{0\}$, by
\begin{align*}
\Psi_{\alpha}&:=\left\lbrace q\in C(\Gamma): \text{(\ref{ConsTBP}) and (\ref{TBP}) hold}\right\rbrace,
\end{align*}
are nonempty and have nonempty intersection over any finite set of indexes $I\subset \R\backslash\{0\}$. Since each such set $\Psi_{\alpha}$ is closed and convex, and thus weakly closed, the finite intersection property of the weakly compact set $C(\Gamma)$ implies that
\begin{align*}
\Psi:=\bigcap_{\alpha >0}\Psi_{\alpha}(t,x) \neq \emptyset.
\end{align*}
Let now $\tilde{q}\in \Psi$. Define a (signed) measure on $(\R^D\backslash\{0\},\Sigma_{z})$ by setting, for every $A\in \Sigma_{z}$,
\begin{align*}
\overline{\nu}_{z}(A)=\int_{A}\tilde{q}(y)\nu_{z}(dy),
\end{align*}
which is well defined since $\tilde{q}$ is Borel measurable (but not necessarily $\Sigma_z$ measurable). Note that $\overline{\nu}_{z}\ll\nu_{z}$. Then, for every $\alpha\in\R\backslash\{0\}$,
\begin{equation}\label{nuU}
\begin{aligned}
\overline{\nu}_{z}(A^{\alpha}_z)
& = g(z_{\alpha})\indic{\alpha>0}-g(z_{\alpha})\indic{\alpha<0} \\
& = \int_{0}^{\infty}\Gamma_+\left(\nu(z^+_{\alpha}>s)\right)ds
	-\int_{0}^{\infty}\Gamma_-\left(\nu(z^-_{\alpha}>s)\right)ds\\
& = \Gamma_+(\nu(A^{\alpha}_{z}))\indic{\alpha>0}
	-\Gamma_-(\nu(A^{\alpha}_{z}))\indic{\alpha<0},
\end{aligned}
\end{equation}
which implies that the value of $\overline{\nu}_z$ on the sets $A^{\alpha}_z$, $\alpha\in \R\backslash\{0\}$ is independent on the choice of $\tilde{q}\in \Psi$. By the monotone class theorem, the value of $\overline{\nu}_{z}$ on any set $A\in\Sigma_z$ must then be independent on the choice of $\tilde{q}$. Next, set
\begin{align*}
\hat{\psi}_t(\omega,y):=\frac{d\overline{\nu}_{z}}{d\nu_{z}}(y).
\end{align*}
Since, by definition of Radon-Nykodim derivative,
\begin{align*}
\int_A\hat{\psi}_t(\omega,y)\nu(dy)=\int_A\tilde{q}(y)\nu(dy)
\end{align*}
for every $A\in\Sigma_z$, it must be the case that $\hat{\psi}\in C(\Gamma)$. Furthermore, by the $\overline{\nu}_{z}$ a.e. uniqueness of the Radon-Nikodym derivative, if $q\in L^2(\nu)$ is $\Sigma_{z}$-measurable and it satisfies (\ref{ConsTBP}) for every $\alpha>0$, then $q(y)=\hat{\psi}_t(\omega,y)$ for $\nu_{z}$-a.a. $y\in\R^D\backslash\{0\}$. $\blacksquare$

\vspace{1mm}

\noindent \textit{\underline{Step 2}. Fix $(t,\omega)\in[0,T]\times\Omega$, and suppose again $z:=Z_t(\omega,\cdot)$. Then, the function $\hat{\psi}_t(\omega,\cdot)$ defined in step 1 satisfies
\begin{align}\label{hatovereq}
\hat{\psi}_t(\omega,y)=\overline{\psi}_t(\omega,y)
\end{align} 
for $\nu$-a.a. $y\in\R^D\backslash\{0\}$. In particular, this implies, for every $(t,\omega)\in [0,T]\times \Omega$,
\begin{align*}
g(Z_t(\omega,\cdot))
=\int_{\R^D\backslash\{0\}}Z_t(\omega,y)\hat{\psi}_t(\omega,y)\nu(dy)
=\int_{\R^D\backslash\{0\}}Z_t(\omega,y)\overline{\psi}_t(\omega,y)\nu(dy).
\end{align*}
}

\vspace{1mm}

\noindent Proof of step 2.

Fix $(t,\omega)\in[0,T]\times\Omega$, and suppose $z:=Z_t(\omega,\cdot)$. Because $\overline{\psi}_t(\omega,\cdot)$ is really a function of $z$, it must be $\Sigma_{z}$ measurable. To show (\ref{hatovereq}), and based on step 1, we need to show that $\overline{\psi}_t(\omega,\cdot)$ satisfies (\ref{ConsTBP}) for every $\alpha>0$.

To do so, define measures $\overline{\mu}_{z}$ and $\mu_{z}$ on $(\R\backslash\{0\},\mathcal{B}(\R\backslash\{0\})$ as the pushforwards of $\overline{\nu}_z$ and $\nu_z$ under the transformation $z:\R^D\backslash\{0\}\rightarrow \R$, i.e. set, for every $\alpha>0$,
\begin{align*}
\overline{\mu}_{z}([\alpha,\infty))
&=\overline{\nu}_{z}(A^{\alpha}), \
\overline{\mu}_{z}((-\infty,-\alpha])
=\overline{\nu}_{z}(A^{-\alpha}), \\
\mu_{z}([\alpha,\infty))
&=\nu_{z}(A^{\alpha}), \
\mu_{z}((-\infty,-\alpha])
=\nu_{z}(A^{-\alpha}),
\end{align*}
where $\overline{\nu}_{z}$ is as in step 1 of the proof. Then, for $\mu_{z}$-a.a. $\alpha\in\R\backslash\{0\}$,
the ball ratio limit representation\footnote{For the ball ratio limit representation of the Radon-Nykodim derivative see \cite{Bogachev}, Theorem 5.5.8.} of the Radon-Nykodim derivative on $\R$ implies that
\begin{align*}
\frac{d\overline{\mu}_{z}}{d\mu_{z}}(\alpha)
=\lim_{\varepsilon\downarrow 0}
\frac{\Gamma_+(\nu(A_z^{\alpha-\varepsilon}))-
		\Gamma_+(\nu(A_z^{\alpha+\varepsilon}))}
	{\nu(A_z^{\alpha-\varepsilon})-
		\nu(A_z^{\alpha+\varepsilon})}
=\Gamma_+'(\nu(A_z^{\alpha})),
\end{align*}
if $\alpha>0$ and
\begin{align*}
\frac{d\overline{\mu}_{z}}{d\mu_{z}}(\alpha)
=-\lim_{\varepsilon\downarrow 0}
\frac{\Gamma_-(\nu(A_z^{\alpha+\varepsilon}))-
	\Gamma_+(\nu(A_z^{\alpha-\varepsilon}))}
{\nu(A_z^{\alpha+\varepsilon})-
	\nu(A_z^{\alpha-\varepsilon})}
=-\Gamma_-'(\nu(A_z^{\alpha})),
\end{align*}
if $\alpha<0$. 
%Now, define, for $y\R^D\backslash\{0\}$,
%\begin{align*}
%\tilde{\psi}^U(t,x,y)
%& := \Gamma_+'(\nu(A^{z(t,x,y)}_{t,x})\indic{z(t,x,y)>0}
%	 -\Gamma_-'(\nu(A^{z(t,x,y)}_{t,x})\indic{z(t,x,y)<0}
%\end{align*}
%and
Next note that for every $B\subset \mathcal{B}(\R\backslash\{0\})$ and $\mathcal{B}(\R\backslash\{0\})$-measurable function $\theta$,
\begin{align}\label{Lebesgue}
\int_{B}\theta(p)\mu_{z}(dp)
=\int_{z^{-1}(B)}\theta(z(y))\nu_{z}(dy),
\end{align}
which holds since $\mu_{z}$ is the pushforward of $\nu_{z}$ under $z$. Then, for every $\alpha>0$,
\begin{align*}
\int_{A^{\alpha}}\overline{\psi}_t(\omega,y)\nu(dy)
&=\int_{\alpha}^{\infty}\Gamma'_+(\nu(A^p))\mu_{z}(dp)\\
&=\int_{\alpha}^{\infty}
 \frac{d\overline{\mu}_z}{d\mu_z}(p)
 \mu_z(dp)\\
&=\Gamma_+(\nu(A^{\alpha})),
\end{align*}
and, for every $\alpha<0$,
\begin{align*}
\int_{A^{\alpha}}\overline{\psi}_t(\omega,y)\nu(dy)
&=-\int_{-\infty}^{\alpha}\Gamma'_-(\nu(A^p))\mu_{z}(dp)\\
&=\int_{\alpha}^{\infty}
 \frac{d\overline{\mu}_z}{d\mu_z}(p)
 \mu_z(dp)\\
&=-\Gamma_-(\nu(A^{\alpha})),
\end{align*}
That is, $\overline{\psi}_t(\omega,\cdot)$ satisfies (\ref{ConsTBP}), so that $\hat{\psi}_t(\omega)=\overline{\psi}$ $\nu_{t,x}$-a.e., which in turn implies that $\overline{\psi}_t(\omega,\cdot)\in C(\Gamma)$ and satisfies (\ref{TBP}).$\blacksquare$

\vspace{1mm}

\noindent\textit{\underline{Step 3}. Conclusion.}

\vspace{1mm}

\noindent Proof of step 3.

By Proposition \ref{remSMM} and since $\overline{Z}$ and $\underline{Z}$ are predictable, $\overline{\psi}$ and $\underline{\psi}$ are predictable selectors of $\partial g(\overline{Z}_t)$ and $\partial g(\underline{Z}_t)$ respectively for $t\in [0,T]$, and the stochastic integrals
\begin{align*}
\overline{M}_t = \int_{[0,t]\times \R^D\backslash\{0\}}\overline{\psi}_s(y)\tilde{N}(ds,dy), \
\underline{M}_t = \int_{[0,t]\times \R^D\backslash\{0\}}\underline{\psi}_s(y)\tilde{N}(ds,dy)
\end{align*}
are well defined. By Lemma A.24 in \cite{LaevenStadje}, the associated measures $\overline{\Q}$ and $\underline{\Q}$ are also well defined, and the proof is complete.
\end{proof}

\section{The Case of Deterministic Level Sets of the Control Processes}\label{Monotone}
\subsection{The Extreme Measures of Monotone Claims}
If $D=1$ and the claim's payoff is monotonic and Lipschitz continuous, then the processes $\overline{\psi}$ and $\underline{\psi}$ in Theorem \ref{SMMformula} are deterministic, and so they can be fully specified.  We begin by showing the following result.
\begin{proposition}\label{ThMartingaleMonotonic}
Consider the setup of Theorem \ref{SMMformula}, and suppose that $D=1$. Suppose that for some $(t,\omega)\in [0,T]\times\Omega$, $\overline{Z}_t(\omega,\cdot)$ is non-decreasing and that, for $y\in\R\backslash\{0\}$,
\begin{align*}
\sign(\overline{Z}(y))=\sign(y).
\end{align*}
Then, the process $\overline{\psi}$ defined in Theorem \ref{SMMformula} is deterministic and time independent, and is given by the function $\overline{\psi}_{\Gamma}$ specified in (\ref{psifun}), i.e. for every $y\in\R\backslash\{0\}$, $t,\in [0,T]$
\begin{align*}
\overline{\psi}_t(\omega,y) = \overline{\psi}_{\Gamma}(y)
	& := \Gamma_+'\left(\nu([y,\infty))\right)\indic{y>0}
    		-\Gamma_-'\left(\nu((-\infty,y]))\right)\indic{y<0}.
\end{align*}
If $\overline{Z}\in L^p(\nu)$ is non-increasing, then, for every $y\in \R\backslash\{0\}$, $\overline{\psi}$ is given by the function $\underline{\psi}_{\Gamma}$ specified in (\ref{psifun}), i.e. for every $y\in\R\backslash\{0\}$, $t,\in [0,T]$,
\begin{align*}
\overline{\psi}_t(\omega,y) = \underline{\psi}_{\Gamma}(y)
& :=  -\Gamma_-'\left(\nu([y,\infty))\right)\indic{y>0}
	+\Gamma_+'\left(\nu((-\infty,y]))\right)\indic{y<0}.
\end{align*}
Conversely, the process $\underline{\psi}$ is given by $\underline{\psi}_{\Gamma}$ if $\underline{Z}$ is non-decreasing, and by $\overline{\psi}_{\Gamma}$ if $\underline{Z}$ is non-increasing.
\end{proposition}
\begin{proof}
This result follows by noting that if $\overline{Z}_t(\omega,\cdot)$ is non-decreasing and, for $y\neq 0$, 
\begin{align*}
\sign(\overline{Z}(y))=\sign(y),
\end{align*}
then, for $y\neq 0$,
\begin{equation*}
\indic{y>0}=\indic{z(y)>0},\\
\indic{y<0}=\indic{z(y)<0},
\end{equation*}
and
\begin{align}\label{Z_monotone}
\{\overline{Z}_t(\omega,\cdot)>\overline{Z}(\omega,y)\}=
\begin{cases}
[y,\infty), \text{ if } y>0,\\
(-\infty,y], \text{ if } y<0.
\end{cases}
\end{align}
Plugging (\ref{Z_monotone}) into (\ref{psiformula}) yields the result. The other cases are similar.
\end{proof}

In order to apply Proposition \ref{ThMartingaleMonotonic} we need to be able to fully specify the process $Z$ in \ref{BSDETh}. This can be done if any of the assumptions in Theorem \ref{ThBSDEPIDE} are satisfied.

\begin{theorem}\label{MonotoneMeasures} Suppose a claim pays off $C=f(X_T)\in L^2$ at time $t$, where $f:\R\rightarrow \R$ is Borel measurable, non-decreasing and Lipschitz continuous. Let $u$ and $\ell$ be the deterministic functions corresponding to the valuation bounds $U$ and $L$ for $C$. Suppose $u\in C^{0,1}([0,T]\times \R)$ or the probability law of $X^{t,x}_s$ for $s\geq t\geq 0$ is absolutely continuous with respect to the Lebesgue measure. Then, there is an upper extreme measure $\overline{\Q}(\Gamma)$ for $C$ such that $X$ is a L\'evy process under $\overline{\Q}(\Gamma)$ with L\'evy measures $\overline{\nu}$ defined by
\begin{align}\label{Upper_nu}
\overline{\nu}(A)&=\nu(A)+\Gamma_+(\nu(A\cap (0,\infty)))-\Gamma_-(\nu(A\cap (-\infty,0))), A\in \mathcal{B}(\R\backslash\{0\}).
\end{align}
Similarly, if $\ell\in C^{0,1}([0,T]\times \R)$ or the probability law of $X^{t,x}_s$ for $s> t\geq 0$ is absolutely continuous with respect to the Lebesgue measure, there is a lower extreme measure $\underline{\Q}(\Gamma)$ for $C$ such that
$X$ is a L\'evy process under $\underline{\Q}(\Gamma)$ with L\'evy measures $\underline{\nu}$ defined by
\begin{align}\label{Lower_nu}
\underline{\nu}(A)&=\nu(A)-\Gamma_-(\nu(A\cap (0,\infty)))+\Gamma_+(\nu(A\cap (-\infty,0))), \ A\in \mathcal{B}(\R\backslash\{0\}).
\end{align}
Finally, similar results hold if $f$ is non-increasing.
\end{theorem}
\begin{proof}
Let $f$ be as in the Theorem's statement and let $(U,\overline{Z})$ be the solution of the BSDE (\ref{BSDE}). By Theorem \ref{PIDE}, if $u\in C^{0,1}([0,T]\times \R)$ or the probability law of $X^{t,x}_s$ for $s> t\geq 0$ is absolutely continuous with respect to the Lebesgue measure, then $\overline{Z}$ satisfies
\begin{align*}
\overline{Z}_t(y)&=u(t,X_{t-}+y)-u(t,X_{t-}).
\end{align*}
Next note that, for every $x',x\in\R$, $x'>x$ implies $X^{t,x'}_s\geq X^{t,x}_s$ $\Q$-a.s. for every $s>t\geq 0$, and so, assuming $f$ non-decreasing, $f(X^{t,x'}_T)\geq f(X^{t,x}_T)$ $\Q$-a.s. Hence, by the Comparison Principle (\cite{Delong}, Theorem 3.2.1), $u(t,x')\geq u(t,x)$. Therefore, $u(t,x,\cdot)$ is non-decreasing for every $(t,x)\in [0,T]\times\R^D$ if $f$ is non-decreasing. But then $\overline{Z}_t(\omega,\cdot)$ is also non-decreasing and satisfies $\sign(\overline{Z}(y))=\sign(y)$. Then, by Proposition \ref{ThMartingaleMonotonic}, the process $\overline{\psi}$ in Theorem \ref{SMMformula} is satisfies for every $y\in\R\backslash\{0\}$, $t\in[0,T]$,
\begin{align*}
\overline{\psi}_t(\omega,y)
& = \Gamma_+'\left(\nu([y,\infty))\right)\indic{y>0}
	-\Gamma_-'\left(\nu((-\infty,y]))\right)\indic{y<0}
	=\overline{\psi}_{\Gamma}(y).
\end{align*}
Similarly, the process $\underline{\psi}$ in Theorem \ref{SMMformula} satisfies
\begin{align*}
\underline{\psi}_t(\omega,y)
& = -\Gamma_-'\left(\nu([y,\infty))\right)\indic{y>0}
	+\Gamma_+'\left(\nu((-\infty,y]))\right)\indic{y<0}
	=\underline{\psi}_{\Gamma}(y).
\end{align*}
Finally, by Girsanov theorem (specifically, Theorem 3.17 and Theorem 5.19 in \cite{JacodShiryaev}),
\begin{align*}
\overline{\nu}(dy)&=(1+\overline{\psi}(y))\nu(dy),\
\underline{\nu}(dy)=(1+\underline{\psi}(y))\nu(dy),
\end{align*}
which yield (\ref{Upper_nu}) and (\ref{Lower_nu}) respectively.
\end{proof}

\begin{remark}\label{ComRem} Under the assumptions of Theorem \ref{MonotoneMeasures}, and as in static Conic Finance, the valuation bounds are linear over every pairs of claims $C^1=f^1(X_T)$ and $C^2=f^2(X_T)$ where $f^1$ and $f^2$ are both non-decreasing or both non-increasing in $X_T$. It is not true, however, that valuation bounds are linear over any two pair of comonotone claims. Indeed, if this was the case,  there would exist by Remark \ref{LebCont} a measure $\Q^*\in \mathcal{M}$, such that the upper valuations of the claims $C^1$ and $-C^2$, where $C^1$ and $C^2$ are bounded and, say, non-decreasing, are conditional expectations under $\Q^*$. But then, since $\Q^*\in\mathcal{M}$, there is $\psi^*$ such that
\begin{align*}
\E^{\Q}\left[\left.\frac{d\Q^*}{d\Q}
	\right\rvert\F_t\right]=\mathcal{E}(M^*)_t, \
M^*_t =\int_{[0,t]\times \R\backslash\{0\}}\psi^*_s(y)\nu(dy),
\end{align*}
for every $t\in [0,T]$. Thus, $g$ is additive over $\overline{Z}^1_T$ and $\overline{Z}^2_T$, where
\begin{align*}
\overline{Z}^1_T(y)=f^1(X_T+y)-f^1(X_T), \
\overline{Z}^2_T(y)=-f^2(X_T+y)+f^2(X_T).
\end{align*}
However, this cannot be true because $\overline{Z}^1_T$ and $\overline{Z}^2_T$ are not comonotonic and $g$ is a Choquet integral and so it is not linear over functions that are not comonotonic.
\end{remark}

\begin{remark}\label{LawInvariance} Theorem \ref{MonotoneMeasures} also shows that the valuation bounds of monotonic payoff functions $f^1$ and $f^2$ are the same if the law of $f^1(X_T)$ is the same as that of $f^2(X_T)$. Indeed, in this case, the Levy measures of $X$ under the upper extreme measures implied by each payoff are the same, and similarly for the lower ones. This property is inherited from the law invariance of static Conic Finance valuations (see \cite{kusuoka}). However, the continuous time Conic Finance bounds are not law invariant in general: the processes $\overline{\psi}_t$ and $\underline{\psi}_t$ defined in Theorem \ref{SMMformula} depend on time and are not deterministic when the payoff function is not monotonic.
% and so the equivalence of the valuation bounds requires equivalence of the processes $\overline{\psi}_t$ associated to each payoff for \textit{all } $t\in[0,T]$, and similarly for $\underline{\psi}_t$.
\end{remark}

\subsection{The Densities of the Extreme Measures of Monotone Claims}

\begin{figure*}
        \centering
        \begin{subfigure}[b]{0.45\textwidth}
            \centering
            \includegraphics[width=\textwidth]
            {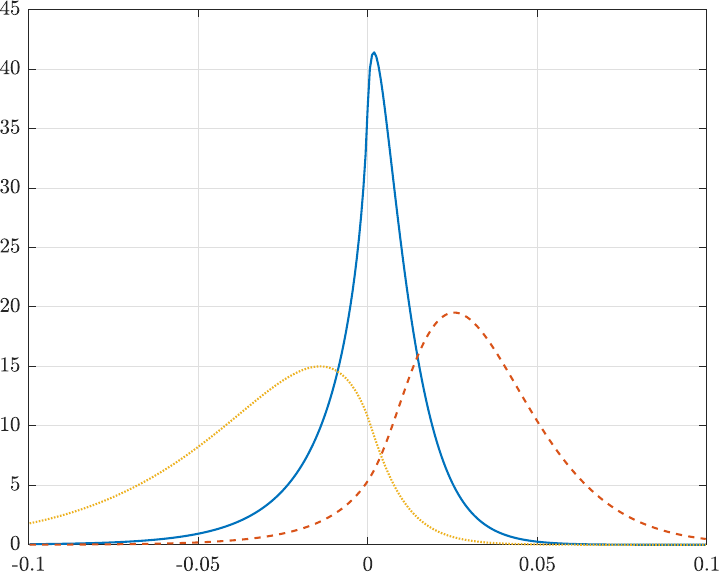}
            \caption{}\label{PDensity}
        \end{subfigure}
		\begin{subfigure}[b]{0.45\textwidth}
            \centering
            \includegraphics[width=\textwidth]
            {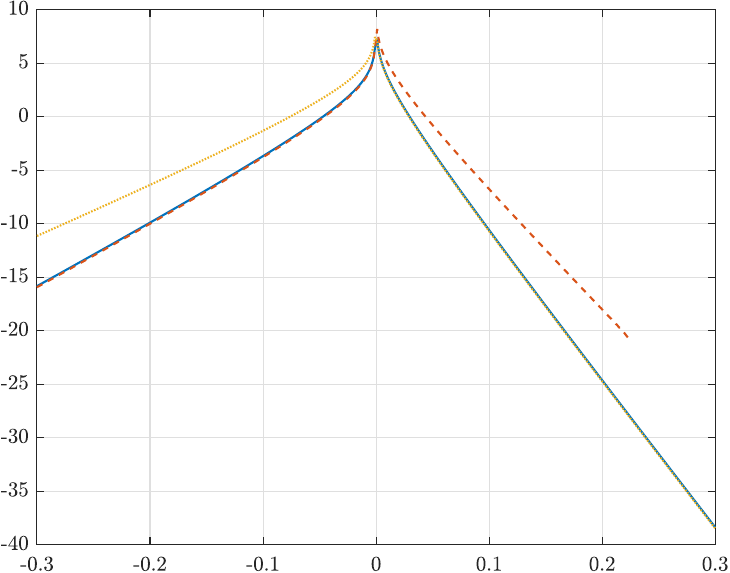}
            \caption{}\label{LDensity}
        \end{subfigure}
%		\centering
%        \begin{subfigure}[b]{0.47\textwidth}
%            \centering
%            \includegraphics[width=\textwidth]
%            {Psi_U-eps-converted-to.pdf}
%            \caption{}
%        \end{subfigure}
%		\begin{subfigure}[b]{0.47\textwidth}
%            \centering
%            \includegraphics[width=\textwidth]
%            {Psi_L-eps-converted-to.pdf}  
%            \caption{}
%        \end{subfigure}
	\caption{(A): plot of the probability densities of $X$ under $\Q$ (solid), $\overline{\Q}(\Lambda)$ (dashed) and $\underline{\Q}(\Lambda)$ (dotted). (B): plot of the log L\'evy densities.}\label{Density}
\end{figure*}

The characteristic exponents under the measures $\overline{\Q}(\Gamma)$ and $\underline{\Q}(\Gamma)$ in Theorem \ref{MonotoneMeasures} of the process $X$ can be numerically computed for $t\in[0,T]$ based on the L\'evy-Kintchine formula as, respectively,
\begin{align*}
\E^{\overline{\Q}(\Gamma)}\left[e^{i\theta X_t}|X_0=x\right]
&=e^{t\left(i\theta \mathtt{d}+\int_{\R\backslash\{0\}}(e^{i\theta y}-1)(1+\overline{\psi}_{\Gamma}(y))\nu(dy)\right)}\\
\E^{\underline{\Q}(\Gamma)}\left[e^{i\theta X_t}|X_0=x\right]
&=e^{t\left(i\theta\mathtt{d}+\int_{\R\backslash\{0\}}(e^{i\theta y}-1)(1+\underline{\psi}_{\Gamma}(y))\nu(dy)\right)}.
\end{align*}
The corresponding probability densities can then be obtained via Fourier inversion. For instance, the densities of $X_1$ under $\Q$, $\overline{\Q}(\Lambda)$ and $\underline{\Q}(\Lambda)$, are plotted in Figure \ref{Density}(A), under the assumption that the measure distortions $\Gamma_+$ and $\Gamma_-$ are the distortions $\Lambda_+$ and $\Lambda_-$ specified by equation (\ref{MeasureDistortion}), with parameters 
\begin{align*}
(c,\gamma,a,b)=(0.01,0.25,100,1),
\end{align*}
and that, under $\Q$, $X$ is a BG process with drift $\mathtt{d}$. This means that the the compensator of $N$ satisfies 
\begin{align}\label{BGdensity}
\nu(dy)
 = \left(\frac{c_pe^{-y/b_p}}{y}\indic{y>0}
	+\frac{c_ne^{-|y|/b_p}}{|y|}\indic{y<0}\right)dy
\end{align}
for $y\in\R\setminus\{0\}$ and positive scale parameters $b_p$ and $b_n$ and speed parameters $c_p$ and $c_n$. See \cite{MadanEntropy} for an interpretation of these parameters in terms of the structure of market and limit orders. The assumption that $\{Y_0e^{X_t}\}_{t\geq 0}$ is a martingale under $\Q$ then implies 
\begin{align}\label{BGX}
\mathtt{d} = -c_p\log(1-b_p)-c_n\log(1+b_n).
\end{align}
In Figure \ref{Density}(A), we set 
\begin{align*}
(b_p,c_p,b_n,c_n) = (0.0075, 1.5592, 0.0181, 0.6308),
\end{align*}
obtained by estimation to SPY prices observed between 2 January 2020 and 31 December 2020. The probability density shown in Figure \ref{Density}(A) is for a specific tenor (here set to 1 month).  The log L\'evy density of $X$ under $\Q$, $\overline{\Q}(\Lambda)$ and $\underline{\Q}(\Lambda)$ is plotted in Figure \ref{Density} (B).

% Furthermore, as the time $t$ decreases to $0$, both the upper and lower densities tend to look alike, and so one can expect the drifts of the upper valuation to be lower than that of the lower one. 

\section{Empirical Study I: Calibration of Distortions to Option Prices}\label{Calibration}

\subsection{Calibration to Options Bid-Ask Spreads}
Practically, $\Gamma_+$ and $\Gamma_-$ are specified by a parametric family of measure distortions, and the parameters can be calibrated by matching model's upper and lower options prices to their respective market's ask and bid prices. Such calibration exercise is performed in this section and the resulting laws of $X$ under the upper and lower market implied extreme measures is observed.

To do so, we assume \textbf{in this section} a \textbf{nonzero risk free rate} $r$, and so the pricing formulas are adjusted to consider that. Also, we assume again that $X$ is, under $\Q$, a BG process with parameters $(b_p,c_p,b_n,c_n)$ and drift
\begin{align}\label{BG_Q}
\mathtt{d} = r-c_p\log(1-b_p)-c_n\log(1+b_n),
\end{align}
Hence, $Y=\{Y_0e^{-rt+X_t}\}_{t\geq 0}$ is a $\Q$-martingale for any constant $Y_0$. In this study, we take $Y_0$ as the daily closing mid price of the security $Y$.\footnote{Alternatively, $Y_0$ can be a parameter to be calibrated.}

\begin{proposition}\label{OptionsPayoffs} Suppose $X$ is the BG process given by (\ref{BG_Q}). Fix a strike $K\in [0,\infty)$, and let $\overline{\Q}(\Gamma)$ and $\underline{\Q}(\Gamma)$ be the extreme measures in Theorem \ref{MonotoneMeasures}. Then, for every $t\in[0,T]$,
\begin{align}
\esssup_{\Q^{\psi}\in \mathcal{M}}\E^{\Q^{\psi}}[e^{-r(T-t)}\left(Y_0e^{X_T}-K\right)^+|\F_t]&=\E_t^{\overline{\Q}(\Gamma)}\left[e^{-r(T-t)}\left(Y_0e^{X_T}-K\right)^+\right], \label{OptionPricingRFI}\\
\essinf_{\Q^{\psi}\in \mathcal{M}}\E^{\Q^{\psi}}[e^{-r(T-t)}\left(K-Y_0e^{X_T}\right)^+|\F_t]&=\E_t^{\overline{\Q}(\Gamma)}\left[e^{-r(T-t)}\left(K-Y_0e^{X_T}\right)^+\right],
\label{OptionPricingRFII}
\end{align}
and, similarly 
\begin{align}
\essinf_{\Q^{\psi}\in \mathcal{M}}\E^{\Q^{\psi}}[e^{-r(T-t)}\left(Y_0e^{X_T}-K\right)^+|\F_t]&=\E_t^{\underline{\Q}(\Gamma)}\left[e^{-r(T-t)}\left(Y_0e^{X_T}-K\right)^+\right], \label{OptionPricingRF_QLI}\\
\esssup_{\Q^{\psi}\in \mathcal{M}}\E^{\Q^{\psi}}[e^{-r(T-t)}\left(K-Y_0e^{X_T}\right)^+|\F_t]&=\E_t^{\underline{\Q}(\Gamma)}\left[e^{-r(T-t)}\left(K-Y_0e^{X_T}\right)^+\right].
\label{OptionPricingRF_QLII}
\end{align}
\end{proposition}
\begin{proof}
The existence of $\overline{\Q}(\Gamma)$ and $\underline{\Q}(\Gamma)$ is guaranteed by Theorem \ref{MonotoneMeasures}, since the probability distribution of $X_s^{t,x}$ for $s\geq t\geq 0$ is absolutely continuous when $X$ is a BG process. The payoff functions of calls and puts are not Lipschitz continuous, but, for every $\Q^{\psi}\in\mathcal{M}$,
\begin{align*}
\lim_{n\rightarrow \infty} \E^{\Q^{\psi}}\left[
\max\left(\left(Y_0e^{X_T}-K\right)^+,n\right)|\F_t\right]
=\E^{\Q^{\psi}}\left[\left(Y_0e^{X_T}-K\right)^+|\F_t\right]
\end{align*}
by the monotone convergence theorem, and so the proof of (\ref{OptionPricingRFI}) follows from 
\begin{align*}
\E_t^{\Q^{\psi}}\left[
\max\left(\left(Y_0e^{X_T}-K\right)^+,n\right)\right]
\leq 
\E_t^{\overline{\Q}(\Gamma)}\left[
\max\left(\left(Y_0e^{X_T}-K\right)^+,n\right)\right]
\end{align*}
for every $n\in\N$, which holds by Theorem \ref{MonotoneMeasures}. The proofs of the remaining equations are similar.
\end{proof}
Based on Proposition \ref{OptionsPayoffs}, the FFT-based method developed in \cite{CarrMadanFFT} can be employed to calibrate BG and measure distortion parameters to OTM\footnote{As usual, the calibration is performed on OTM options as these are more liquid than in the money (ITM) ones.} options.

\subsection{A New Family of Measure Distortions}
One common family of measure distortions is the one introduced in Example \ref{MeasureDistortionEx}. The extreme measures $\overline{\Q}(\Lambda)$ and $\underline{\Q}(\Lambda)$ associated to them via Theorem \ref{MonotoneMeasures} are constructed as in Definition \ref{NonlinearVal} by the functions
\begin{align*}
\overline{\psi}_{\Lambda}(y)&=\frac{ac}{1+\gamma}\left(1-e^{-c\nu([y,\infty))}\right)^{-\frac{\gamma}{1+\gamma}}e^{-c\nu(y,\infty)}\indic{y>0}
-be^{-c\nu((-\infty,y])}\indic{y<0},\\
\underline{\psi}_{\Lambda}(y)&=
be^{-c\nu([y,\infty)}\indic{y>0}
-\frac{ac}{1+\gamma}\left(1-e^{-c\nu((-\infty,y])}\right)^{-\frac{\gamma}{1+\gamma}}e^{-c\nu((-\infty,y])}\indic{y<0}.
\end{align*}
%It is important to note that $\overline{\psi}$ and $\underline{\psi}$ may not be in $L^2(\nu)$ if assumption \ref{MDCondition} is not satisfied. For instance, if $\Gamma_+$ is as in \ref{MeasureDistortion}, then, for $y\geq 0$,
%
%and if $\gamma\geq 1$ and $\nu$ an exponential integral (as e.g. in \cite{MadanSeneta} or \cite{KuchlerTappe}), $\overline{\psi}\notin L^2(\nu)$. Based on the properties of the set $C(\Gamma)$, one could extend the theory of spectral martingale measures to the general Banach space $L^{p}(\nu)$ with $p\geq 2$. This would require however a definition of stochastic integral for processes that are not square integrable (as in \cite{LpStochInt}) and a generalization of the theory of spectral risk measures, so a further investigation of this issue is left for future research. Instead, upper and lower spectral martingale measures for monotonic claims in dimension $D=1$ are described next, and the requirement that $\overline{\psi}$ and $\underline{\psi}$ be square integrable is enforced in our estimations.

Another possibility is to define measure distortions $\Upsilon_+$ and $\Upsilon_-$ such that, under the extreme measures $\overline{\Q}(\Upsilon)$ and $\underline{\Q}(\Upsilon)$ associated to them via Theorem \ref{MonotoneMeasures}, the process $X$ is a BG process with parameters 
\begin{align*}
(\overline{b}_p,\overline{c}_p,\overline{b}_n,\overline{c}_n), \
(\underline{b}_p,\underline{c}_p,\underline{b}_n,\underline{c}_n),
\end{align*}
respectively, and given that it is a BG process under $\Q$. For this to be the case, and based on Girsanov's Theorem (Theorems 3.17 and 5.19 in \cite{JacodShiryaev}), the respective functions $\overline{\psi}_{\Upsilon}$ and $\underline{\psi}_{\Upsilon}$ defined in Proposition \ref{ThMartingaleMonotonic} must satisfy
\begin{align}
\overline{\psi}_{\Upsilon}(y)&=\frac{\overline{\kappa}(y)}{\kappa(y)}-1, \label{upcall}\\
\underline{\psi}_{\Upsilon}(y)&=\frac{\underline{\kappa}(y)}{\kappa(y)}-1 \label{downcall}
\end{align}
where $\overline{\kappa}$, $\kappa$ and $\underline{\kappa}$ are the BG L\'evy densities under $\overline{\Q}(\Upsilon)$, $\Q$ and $\underline{\Q}(\Upsilon)$ respectively. On the other hand, we know that
\begin{align*}
\overline{\psi}_{\Upsilon}(y)
=\Upsilon_+'(\nu([y,\infty))), \
\overline{\psi}_{\Upsilon}(-y)
=-\Upsilon_-'(\nu((-\infty,-y])),
\end{align*}
for every $y>0$, which implies
\begin{align*}
\overline{c}_p\frac{e^{-y/\overline{b}_p}}{y}
-c_p\frac{e^{-y/b_p}}{y}
	&=\Upsilon_+'(\nu([y,\infty)))c_p\frac{e^{-y/b_p}}{y},\\
\overline{c}_n\frac{e^{-y/\overline{b}_n}}{y}
-c_n\frac{e^{-y/b_n}}{y}
	&=\Upsilon_-'(\nu((-\infty,-y]))c_n\frac{e^{-y/b_n}}{y}.
\end{align*}
Integrating both sides of the above equations then yields
\begin{align}\label{BG2BGnu}
\Upsilon_+(\nu([y,\infty)))
	&=\left[\overline{c}_pE_1(y/\overline{b}_p)-c_pE_1(y/b_p)\right], \\ 
\Upsilon_-(\nu((-\infty,-y]))
	&=-\left[c_nE_1(-y/\overline{b}_n)-c_nE_1(-y/b_n)\right], \ 
\end{align}
or, for $x>0$,
\begin{equation}\label{BG2BG}
\begin{aligned}
\Upsilon_+(x)=\frac{1}{\overline{c}_p}E_1[E_1^{-1}(x/c_p)b_p/\overline{b}_p]-x, \\
\Upsilon_-(x)=-\frac{1}{\overline{c}_n}E_1[E_1^{-1}(x/c_n)b_n/\overline{b}_n]+x,
\end{aligned}
\end{equation}
where, for every $\alpha>0$, $E_{\alpha}$ is defined as:
\begin{align*}
E_{\alpha}(x)=\int_x^{\infty}\frac{e^{-t}}{t^{\alpha}}dt.
\end{align*}
Then, assuming $c_p=c_n$, by Theorem \ref{MonotoneMeasures}, the L\'evy measures $\overline{\nu}$ and $\underline{\nu}$ of $X$ under, respectively, $\overline{\Q}(\Upsilon)$ and $\underline{\Q}(\Upsilon)$ satisfy
\begin{align*}
\overline{\nu}([y,\infty))
& = \overline{c}_pE_1(y/\overline{b}_p), \
\overline{\nu}((-\infty,y])
  = \overline{c}_nE_1(y/\overline{b}_n), \\
\underline{\nu}([y,\infty))
&= \overline{c}_p
	E_1\left(y\tfrac{b_n}{b_p\overline{b}_n}\right), \
\underline{\nu}((-\infty,y])
  = \overline{c}_n
  	E_1\left(y\tfrac{b_p}{b_n\overline{b}_p}\right).
\end{align*}
Hence, it must be the case that
\begin{align}\label{UpsCondI}
\underline{b}_p = \tfrac{b_p\overline{b}_n}{b_n}, \
\underline{b}_n = \tfrac{b_n\overline{b}_p}{b_p}.
\end{align}
Finally, since $\overline{\Q}(\Upsilon),\underline{\Q}(\Upsilon)\ll\Q$, the Hellinger distance between the measures $\overline{\nu}$ and $\nu$, and $\underline{\nu}$ and $\nu$ must be finite (Theorem IV.4.39 in \cite{JacodShiryaev}). As shown in \cite{KuchlerTappe}, this is possible if and only if the speed parameters satisfy
\begin{align*}
\overline{c}_p&=c_p=\underline{c}_p, \
\overline{c}_n =c_n=\underline{c}_n,
\end{align*}
which, together with the assumption that $c_p=c_n$, implies
\begin{align}\label{UpsCondII}
\overline{c}_p=c_p=\underline{c}_p = 
\overline{c}_n=c_n=\underline{c}_n,
\end{align}
Condition (\ref{UpsCondI}) is needed to guarantee that buying a long position in an asset is equivalent to selling a short position in it. Condition (\ref{UpsCondII}) is necessary to ensure that $\Upsilon_+$ and $\Upsilon_-$ are measure distortions and that $X$ is a BG process under $\overline{\Q}(\Upsilon)$ and $\underline{\Q}(\Upsilon)$. The next Proposition identifies sufficient conditions for $\Upsilon_+$ and $\Upsilon_-$ to be measure distortions.
\begin{proposition}\label{BG2BGProp} The distortions $\Upsilon_+$ and $\Upsilon_-$ defined by (\ref{BG2BG}) with $\overline{b}_p\geq b_p>\overline{b}_p/2$, $c_p=\overline{c}_p$, $c_n=\overline{c}_n$ and $\overline{b}_n\leq b_n$, are bounded, increasing, concave, satisfy assumption \ref{MDCondition} and, for every $x\geq 0$, $\Upsilon_-(x)\leq x$.
\end{proposition}
\begin{proof}
From (\ref{BG2BGnu}), since, for every $y>0$,
\begin{align*}
\frac{d}{dy}\Upsilon_+(\nu([y,\infty)))
=-\overline{c}_p\frac{e^{-y/\overline{b}_p}}{y}
	+c_p\frac{e^{-y/b_p}}{y},
\end{align*}
and since $\nu([y,\infty))$ is decreasing, $\Upsilon_+$ is increasing in $y$ if and only if $\overline{b}_p>b_p$. In this case, in order for $\Upsilon_+$ to be bounded it is necessary and sufficient to show that
\begin{align*}
\lim_{y\rightarrow 0^+}\Upsilon_+(\nu([y,\infty)))<\infty.
\end{align*}
If $\overline{c}_p\neq c_p$, the above limit is infinity, so $\overline{c}_p=c_p$. Then, using 5.1.11 in \cite{AbramovitzStegun},
\begin{align*}
\lim_{y\rightarrow 0}\Upsilon_+(\nu([y,\infty)))
& = \lim_{y\rightarrow 0} \left[\log\left(\frac{y}{\overline{b}_p}\right)
 	+\sum_{k=1}^{\infty}\frac{(-y/\overline{b}_p)^k}{kk!}
 	-\log\left(\frac{y}{b_p}\right)+\sum_{k=1}^{\infty}\frac{(-y/b_p)^k}{kk!}\right]\\
& = \log\left(\frac{\overline{b}_p}{b_p}\right),
\end{align*}
where we used the fact that the sums inside the square brackets converge absolutely. To prove concavity, note that
\begin{align*}
\Upsilon_+'(x)
&=\frac{e^{-E_1^{-1}(x c_p)b_p/\overline{b}_p}}{E_1^{-1}(x c_p)c_pb_p/\overline{b}_p}
	\frac{E_1^{-1}(x c_p)c_pb_p/\overline{b}_p}{e^{-E_1^{-1}(x c_p)}}-1
=e^{-E_1^{-1}(x c_p)(b_p/\overline{b}_p-1)}-1,
\end{align*}
and, if $\overline{b}_p\geq b_p$,
\begin{align*}
\Upsilon_+''(x)
=c_pe^{-E_1^{-1}(x c_p)(b_p/\overline{b}_p-1)}\frac{E_1^{-1}(x c_p)}{e^{-E_1^{-1}(x c_p)}}
	\left(\frac{b_p}{\overline{b}_p}-1\right)\leq 0.
\end{align*}
As for assumption (\ref{MDCondition}), using the substitution $xc_p=E_1(y)$ and assuming $\overline{b}_p>b_p>\overline{b}_p/2$, we obtain, for every $0<\varepsilon<b_p/\overline{b}_p-1/2$,
\begin{align*}
\lim_{x\rightarrow 0}\frac{E_1[E_1^{-1}(x c_p)b_p/\overline{b}_p]}{x^{1/2+\varepsilon}}
& = \lim_{y\rightarrow \infty}c_p^{1/2+\varepsilon}
		\frac{E_1(yb_p/\overline{b}_p)}{E_1(y)^{1/2+\varepsilon}}
 = \lim_{y\rightarrow \infty}c_p^{1/2+\varepsilon}\frac{\overline{b}_p}{b_p}
  		e^{-y\left(\frac{b_p}{\overline{b}_p}-\frac{1}{2}-\varepsilon\right)}
  		y^{\varepsilon-1/2} 
 = 0,
\end{align*}
which implies \ref{MDCondition}. The proof for $\Upsilon_-$ is similar, and it is also obvious that $\Upsilon_-(x)\leq x$ for every $x>0$.
\end{proof}
\begin{remark}From Proposition \ref{BG2BGProp} we obtain
\begin{align*}
\underline{b}_p &= \tfrac{b_p\overline{b}_n}{b_n}
	\leq b_p \leq \overline{b}_p, \
\underline{b}_n  = \tfrac{b_n\overline{b}_p}{b_p}
	\geq b_n \geq \underline{b}_n,
\end{align*}
which, together with (\ref{UpsCondII}), ensures that $\E^{\underline{\Q}}[X]\leq	\E^{\overline{\Q}}[X]$.
\end{remark}

\begin{remark} Another way to proceed is to start with (\ref{downcall}), and derive the associated distortions  and parameters $\overline{b}_p$ and $\overline{b}_n$ so that there is consistency between buying $C$ and selling $-C$. In general, there is more than one choice of measure distortions, and thus, of the set of measures $\mathcal{M}$, so that the law of $X$ under the corresponding extreme measures belongs to the same family as the one under $\Q$.
\end{remark}

\subsection{Results of Calibration}\label{CalibrationResults}
\begin{figure*}
		\centering
        \begin{subfigure}[b]{0.4\textwidth}
            \centering
            \includegraphics[width=\textwidth]
            {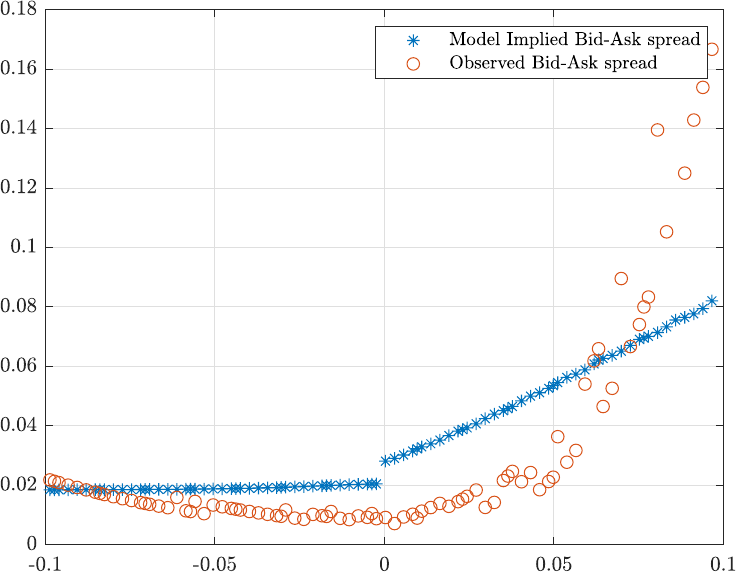}  
            \caption{{}}
        \end{subfigure}
		\centering
        \begin{subfigure}[b]{0.4\textwidth}
            \centering
            \includegraphics[width=\textwidth]
            {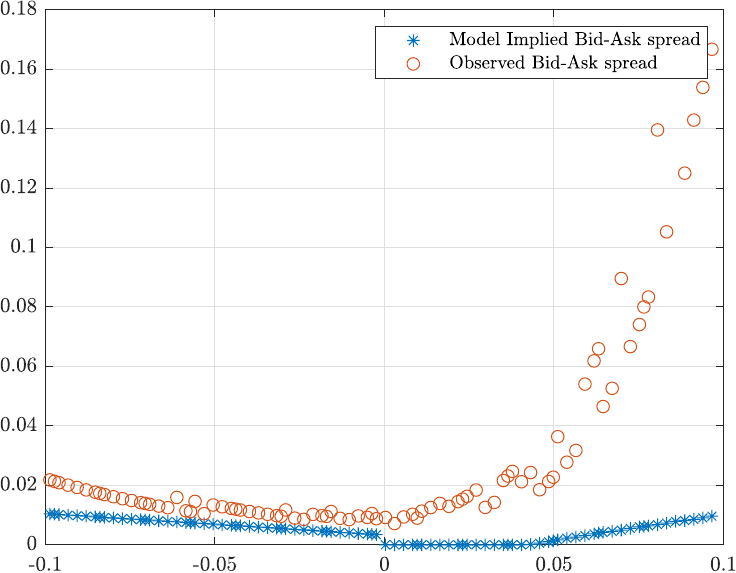}  
            \caption{{}}
        \end{subfigure}
		\caption{Model and market implied relative spreads on options on SPY for distortions $\Lambda_+,\Lambda_-$ (Figure A) and $\Upsilon_+,\Upsilon_-$ (Figure B). Model paramaters were calibratrated to bid and ask prices of OTM options on SPY as of 31 December 2020. Moneyness is represented on the horizontal axis, with negative moneyness referring to OTM put options, and positive moneyness to OTM call options.}\label{RelativeSpread}
\end{figure*}

We considered options on the SPY ETF, and calibrated the measure distortions $\Lambda=(\Lambda_+,\Lambda_-)$ specified in (\ref{MeasureDistortion}), the distortions $\Upsilon=(\Upsilon_+,\Upsilon_-)$ specified in (\ref{BG2BG}), and the BG parameters under $\Q$, to bid and ask prices observed on 31 December 2020 for calls and puts with 1-month expiration. Figures \ref{RelativeSpread}.A and \ref{RelativeSpread}.B show the model and market implied OTM                                                                                                                                                                                                                                                                                                                                                                                                                                                                                                                                                                                                                                                                                                                                                                            options relative bid-ask spreads for the distortions $\Lambda$ and $\Upsilon$ respectively.
%A better fitting is achieved by setting $a=1/c$ and $b=1$ and allowing the parameters $(c,\gamma)$ to differ for bid and ask prices, as shown in figure \ref{Spread}(b). 
%\begin{align*}
%(c^U,\gamma^U,c^L,\gamma^L)
%	& = (3738.8,192.6,12.0,756.6),
%\end{align*}
%where $(c^U,\gamma^U)$ and $(c^L,\gamma^L)$ denote the measure distortion parameters for, respectively, the ask price and the bid price.
The calibrated parameters for the two models are
\begin{align*}
(b_p,c_p,b_n,c_n,c,\gamma,a,b) 
	& = (0.0039,614.5672,0.0979,3.7175,0.0021,0.1996,0.0011,0.0067)\\
(b_p,c_p,b_n,\overline{b}_p,\overline{b}_n)
	& =( 0.0254,7.8699,0.0682,0.0255,0.0681),
\end{align*}
where, $c_n$, $\underline{b}_p$ and $\underline{b}_n$ are obtained for the distortions $\Upsilon$ by conditions (\ref{UpsCondI}) and (\ref{UpsCondII}).
%(0.0019,0.0328,0.0004,0.0065)

For the calibration of the model with measure distortions given by $\Lambda$, and since the characteristic function is not available in closed form in this case, the search of optimal parameters was facilitate by first calibrating the $\Q$-implied BG parameters $(b_p,c_p,b_n,c_n)$ to mid prices of options, and then the parameters $(c,\gamma,a,b)$ to bid and ask prices. In the case of the model based on $\Upsilon$, all parameters $(b_p,c_p,b_n,\overline{b}_p,\overline{b}_n)$ were calibrated directly to bid and ask prices.

\begin{remark}\label{CalibBG2BG}
Because of (\ref{UpsCondII}), the measure distortions $\Gamma_+$ and $\Gamma_-$ specified by (\ref{MeasureDistortion}) outperform those specified by (\ref{BG2BG}) in replicating bid-ask relative spreads. However, calibration on 31 December 2020 was about 10 times faster for the model specified by $\Upsilon_+$ and $\Upsilon_-$.
\end{remark}

\noindent \textbf{Facts}. 
\begin{itemize}
[noitemsep,nolistsep]
\item[i.] Relative bid-ask spreads of options are higher for OTM calls than for OTM puts (see Figure \ref{RelativeSpread});
\item[ii.] Calibrated parameters $b,c,a$ for $\Lambda_+$ and $\Lambda_-$ imply $\tfrac{b}{c}> 3>0.0011=a$.
\end{itemize}

\begin{figure*}
		\centering
        \begin{subfigure}[b]{0.4\textwidth}
            \centering
            \includegraphics[width=\textwidth]
            {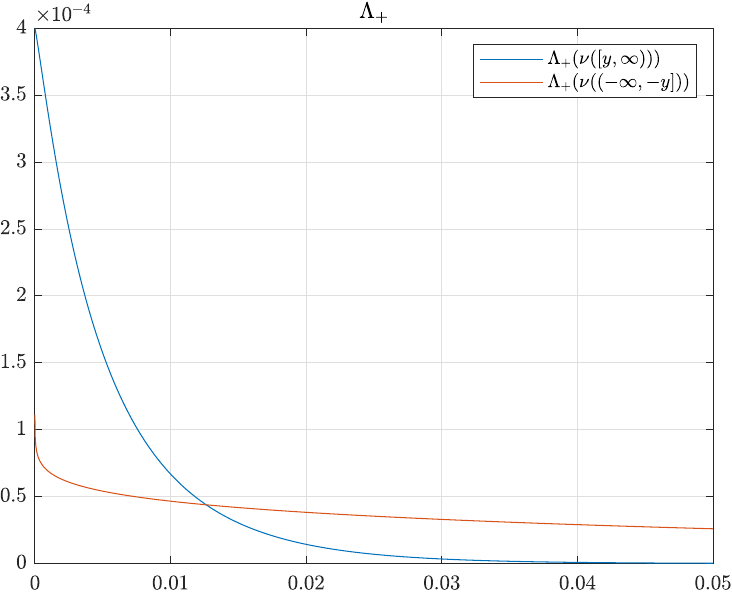}  
            \caption{{}}
        \end{subfigure}
		\centering
        \begin{subfigure}[b]{0.4\textwidth}
            \centering
            \includegraphics[width=\textwidth]
            {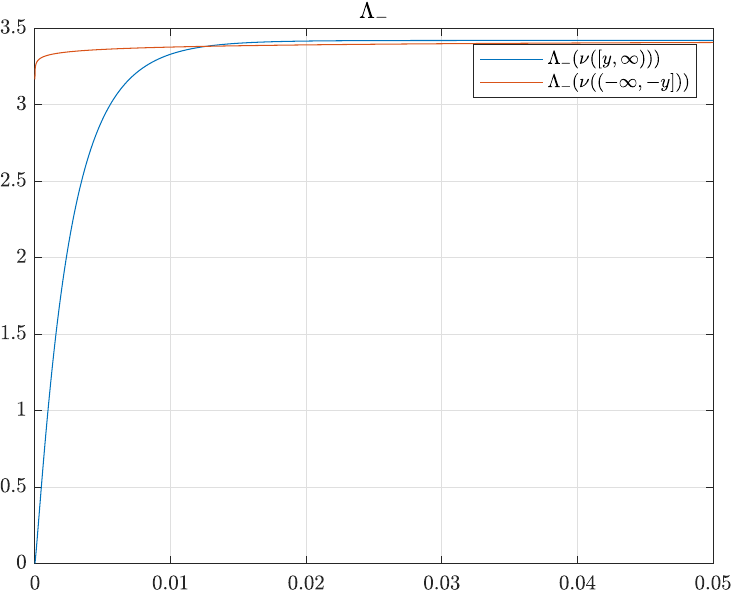}  
            \caption{{}}
        \end{subfigure}
		\caption{The tail measure distortions for $\Lambda_+$ (Figure A) and $\Lambda_-$ (Figure B).}\label{Gamma}
\end{figure*}

\begin{remark}
From fact ii. above, the distortive effect of $\Lambda_-$ on the tail measures is 4 orders of magnitude higher than that of $\Lambda_+$ (Figure \ref{Gamma}). Hence, model implied ask prices of deep OTM calls are close to their observed mid prices, and model bid-ask spreads of calls (and positive delta positions in general) are generated by uncertainty in the loss component of $X$. For puts and negative delta positions, instead, the bid-ask spread is generated by uncertainty in the gain component of $X$. 
\end{remark}
\begin{remark} Because of fact i. (not captured by the distortions $\Upsilon$), the distortive effect of $\Lambda_-$ on the left tail measure is higher than on the right one (see Figure \ref{Gamma}(B)). In fact, denoting by $G=\{G_t\}_{t\geq 0}$ and $L=\{L_t\}_{t\geq 0}$ the gain and loss components of the pure jump part of $X$, we obtain
\begin{align*}
\E^{\Q}[G_T]-\E^{\underline{\Q}}[G_T]
&\approx 0.16, \\
\E^{\overline{\Q}}[L_T]-\E^{\Q}[L_T]
& \approx 0.17.
\end{align*}
\end{remark}

\section{Empirical Study II: Estimation of Distortions From Equity Prices}

\subsection{The Upper and Lower Drifts}\label{Empirical}
As mentioned in the introduction, the third contribution of this paper is to show how Theorem \ref{SMMformula} can be utilized to estimate the shape and size of the set $\mathcal{M}$ of test measures from historical equity prices, and compare such estimate with the one based on the generalized method of moments. Given such purpose, \textbf{in this section} only it will be assumed that the reference probability measure is the \textbf{statistical measure $\P$} of the daily closing mid price process, rather than the measure $\Q$. Specifically, it is meant by this that the probability law of any process estimated based on historical data on daily closing mid prices is an estimate of the process' law under $\P$. Consequently, we assume 
\begin{align*}
X_t=\int_{[0,t]\times\R\backslash\{0\}}xN(dy,ds),
\end{align*}
and we set
\begin{align*}
\mu=\int_{\R\backslash\{0\}}(e^y-1)\nu(dy),
\end{align*}
Then, the asset price process $Y$, identified with the mid-price process of a non-dividend paying stock, satisfies, for every $t\in [0,T]$,
\begin{align*}
Y_te^{-\mu t}=\E^{\P}[Y_0e^{X_T-\mu_T}|\F_t],
\end{align*}
i.e. the discounted process $\{Y_te^{-\mu t}\}_{t\geq 0}$ is a martingale under $\P$, consistently with our assumption that $\P$ is the statistical measure.
%Note that
%\begin{align*}
%Y_t=Y_0+\int_{[0,t]\times \R\backslash\{0\}}Y_{t-}(e^y-1)N(dy,ds),
%\end{align*}
%It was so far assumed that upper and lower valuations for such a claim are given by $U_t=\mathcal{U}[C|\F_t]$ and $L_t=\mathcal{L}[C|\F_t]$. In the formulation with discounting, we let
Then, the results of Theorem \ref{MonotoneMeasures} remain valid, provided they are applied to the processes $\{U_te^{-\overline{\mu} t}\}_{t\geq0}$ and $\{U_te^{-\underline{\mu} t}\}_{t\geq0}$, where 
\begin{align*}
\overline{\mu} = \int_{\R\backslash\{0\}}(e^y-1)\overline{\nu}dy, \ 
\underline{\mu} = \int_{\R\backslash\{0\}}(e^y-1)\underline{\nu}dy,
\end{align*} 
and $\overline{\nu}$ and $\underline{\nu}$ are defined by (\ref{Upper_nu}) and (\ref{Lower_nu}). In fact, a similar argument as in Proposition \ref{OptionsPayoffs}, shows that the supremum and infimum in
\begin{align}\label{DiscountedVal}
U_t:=\sup_{\P^{\psi}\in \mathcal{M}}\E^{\Q^{\psi}}[Y_0e^{X_T}|\F_t], \
L_t:=\inf_{\P^{\psi}\in \mathcal{M}}\E^{\Q^{\psi}}[Y_0e^{X_T}|\F_t]1
\end{align}
are attained for given distortions $\Gamma=(\Gamma_+,\Gamma_-)$ at measures $\overline{\P}(\Gamma)$ and $\underline{\P}(\Gamma)$ analogous to those constructed in Theorem \ref{MonotoneMeasures}. In \ref{DiscountedVal}, $\mathcal{M}$ is defined with respect to $\P$ analogously as in Definition \ref{NonlinearVal}. Then, $U$ and $L$ satisfy, for every $t\in [0,T]$,
\begin{align*}
U_t=e^{\overline{\mu}(T-t)}Y_0e^{X_t}, \ L_t=e^{\underline{\mu}(T-t)}Y_0e^{X_t},
\end{align*}
and so $\{U_te^{-\overline{\mu} t}\}_{t\geq0}$ and $\{L_te^{-\underline{\mu} t}\}_{t\geq0}$ are martingales under $\overline{\P}(\Gamma)$ and $\underline{\P}(\Gamma)$ respectively. Furthermore, an application of Ito's lemma yields
\begin{align*}
U_t&= U_0-\int_{[0,t]\times \R\backslash\{0\}}U_{s-}(e^y-1)\overline{\psi}_{\Gamma}(y)\nu(dy)ds
 +\int_{[0,t]\times \R\backslash\{0\}}U_{s-}(e^y-1)N(dy,ds),\\
L_t&= L_0-\int_{[0,t]\times \R\backslash\{0\}}L_{s-}(e^y-1)\underline{\psi}_{\Gamma}(y)\nu(dy)ds
 +\int_{[0,t]\times \R\backslash\{0\}}L_{s-}(e^y-1)N(dy,ds),
\end{align*}
where $\overline{\psi}_{\Gamma}$ and $\underline{\psi}_{\Gamma}$ are defined as in Proposition \ref{ThMartingaleMonotonic}. Equivalently,
\begin{align*}
\frac{dU_t}{U_{t-}}&=\mu dt-\int_{\R\backslash\{0\}}(e^y-1)\overline{\psi}_{\Gamma}(y)\nu(dy)dt
	+\int_{\R\backslash\{0\}}(e^y-1)\tilde{N}(dy,dt), \\
\frac{dL_t}{L_{t-}}&=\mu dt-\int_{\R\backslash\{0\}}(e^y-1)\underline{\psi}_{\Gamma}(y)\nu(dy)dt
	+\int_{\R\backslash\{0\}}(e^y-1)\tilde{N}(dy,dt).
\end{align*}
where $\tilde{N}$ is a local martingale under $\P$. Taking expectations on both sides of the above equations implies that the upper and lower drifts satisfy
\begin{equation}\label{PricingEquations}
\begin{aligned}
\overline{\mu}dt 
&=\mu dt-\frac{g(U_{t-}(e^{\cdot}-1))}{U_{t-}},\\
\underline{\mu}dt 
&=\mu dt+\frac{g(L_{t-}(e^{\cdot}-1))}{L_{t-}}.
\end{aligned}
\end{equation}
Finally, since the driver function is always nonnegative, we obtain the relation
\begin{align}\label{DriftIneq}
\overline{\mu}\leq \mu \leq \underline{\mu}. 
\end{align}

%An alternative formulation consists in assuming that there exists a risk free asset with risk free rate $r$ and starting with a risk neutral measure $\Q$ instead of $\P$. Exactly the same arguments as above then give
%\begin{align}
%\E^{\Q}\left[\frac{dU_t}{U_{t-}}\right]
%&=r dt-\frac{RC^U_t}{U_{t-}}dt,\label{PricingEquations_r1}\\
%\E^{\Q}\left[\frac{dL_t}{L_{t-}}\right]
%&=r dt+\frac{RC^L_t}{L_{t-}}dt,\label{PricingEquations_r2}
%\end{align}
%and
%\begin{align}\label{DriftIneq_r}
%\int_0^t\E^{\Q}\left[\frac{dU_s}{U_{s-}}\right]\leq rt\leq \int_0^t\E^{\Q}\left[\frac{dL_s}{L_{s-}}\right].
%\end{align}
%This formulation will be further investigated in later sections, when the focus will be moved to general terminal claims.
\subsection{The GMM and DM estimators for the Measure Distortions Parameters}
Given the law of $X$ under $\P$ (estimated from historical observations of daily closing SPY's mid price), one can estimate the size and shape of the set $\mathcal{M}$ by assuming that the measure distortions $\Gamma=(\Gamma_+,\Gamma-)$ belong to a parametric family, such as the one specified in Example \ref{MeasureDistortionEx}. Such parameters are typically estimated using some variations of the generalized method of moments, but, based on Theorems \ref{SMMformula} and \ref{MonotoneMeasures}, it is also possible to obtain via Fourier inversion the probability distribution of $X$ under $\overline{\Q}(\Gamma)$ and $\underline{\Q}(\Gamma)$, and match its tails to those of the empirical distribution of upper and lower valuations. The resulting statistic is  known as the Digital Moment (DM) estimator. In this analysis, we used both estimators to fit measure distortion parameters to observed upper and lower valuations.

Specifically, we assumed in our implementation that $X$ is a BG process with parameters $(b_p,c_p,b_n,c_n)$ obtained through DM estimation from historical observations of daily closing SPY's mid-price. We also assumed that $\Gamma$ is the pairs of distortions $\Lambda=(\Lambda_+,\Lambda_-)$ of Example \ref{MeasureDistortionEx}. We estimated the parameters $(c,\gamma,a,b)$ based on observations on 5-day high and low prices on the SPY and using DM and GMM estimators. Our implementation of the DM estimator is a plain application of the model introduced in \cite{MadanEstimate}, and we refer to that paper for its full description. We outline below the construction of the GMM estimator used. The discrete version of the pricing equations (\ref{PricingEquations}) is given, for discrete times $t=1,...,T$, by
\begin{align*}
\E^{\P}\left[\left.\frac{U_{t+1}-U_t}{U_t}-\int_{\R}(e^y-1)\overline{\psi}_{\Lambda}(y)dy\right\rvert \F_t\right]&=0,\\
\E^{\P}\left[\left.\frac{L_{t+1}-L_t}{L_t}-\int_{\R}(e^y-1)\underline{\psi}_{\Lambda}(y)dy\right\rvert \F_t\right]&=0.
\end{align*}
By iterating expectations, if $h$ is measurable and $\E[|h(U_t)|]<\infty$ and $\E^{\P}[|h(L_t)|]<\infty$,
\begin{align}
\E^{\P}\left[\left.\left(\frac{U_{t+1}-U_t}{U_t}-\int_{\R}(e^y-1)\overline{\psi}_{\Lambda}(y)dy\right)h(U_t)\right\rvert \F_t\right]&=0,
\label{TheoreticalGMM1}\\
\E^{\P}\left[\left.\left(\frac{L_{t+1}-L_t}{L_t}-\int_{\R}(e^y-1)\underline{\psi}_{\Lambda}(y)dy\right)h(L_t)\right\rvert \F_t\right]&=0.
\label{TheoreticalGMM2}
\end{align}

Setting $h(u)=u^k$, $k=1,2,...$ and assuming that \ref{TheoreticalGMM1} and \ref{TheoreticalGMM2} hold, at least, locally, the distortion parameters can be obtained by solving for each $k$ (in our implementation it was assumed $k=1,...,6$),
\begin{align}
\frac{1}{N}\sum_{i=1}^{N}
\left[\left(\frac{U_{t_i+1}-U_{t_i}}{U_{t_i}}-\int_{\R}(e^y-1)\overline{\psi}_{\Lambda}(y)dy\right)U_{t_i}^k\right]&=0\label{GMM1}\\
\frac{1}{N}\sum_{i=1}^N
\left[\left(\frac{L_{t_i+1}-L_t}{L_{t_i}}-\int_{\R}(e^y-1)\underline{\psi}_{\Lambda}(y)dy\right)L_{t_i}^k\right]&=0.\label{GMM2}
\end{align}

The resulting estimators are the GMM estimators.

\subsection{Results of the Estimation}\label{EstimationResults}
We estimated measure distortions parameters via DM and GMM methods for each 5-day non overlapping interval between 1 January 2010 through 31 December 2020. The total of such intervals is 553. Our findings are summarized below.
\vspace{3mm}
\begin{enumerate}
\item \textbf{DM Estimated Distortions are Unbalanced.}

The measure distortion parameters estimated via GMM and DM were quantized into 16 representative points. The five such points with highest representation are shown in Tables \ref{DMParamStats} and \ref{GMMParamStats} for DM and GMM estimates respectively.
\begin{table}[h!]
\centering
\begin{tabular}{| c | c | c | c | c |} 
 \hline
 $c$ & $\gamma$ & $a$ & $b$ & $\tfrac{b}{c}$ \\
 \hline
12.7092 & 0.7689 & 1.1216e-07 & 0.9949 & 0.0783 \\ 
5.8715 & 0.3860 & 8.7573e-06 & 0.9998 & 0.1705 \\ 
9.0998 & 0.4124 & 5.9994e-06 & 0.9982 & 0.1098 \\ 
3.9814 & 0.3558 & 2.0926e-06 & 1.0000 & 0.2514 \\ 
11.3294 & 0.4826 & 9.3874e-06 & 0.9991 & 0.0883 \\ 
 \hline
\end{tabular}\caption{First five quantized points of the DM estimators of the measure distortions parameters $(c,\gamma,b,a)$ for SPY.}
\label{DMParamStats}
\end{table}
\begin{table}[h!]
\centering
\begin{tabular}{| c | c | c | c | c |} 
 \hline
 $c$ & $\gamma$ & $a$ & $b$ & $\tfrac{b}{c}$ \\
 \hline
65.5791 & 0.4550 & 0.0181 & 0.8712 & 0.0133 \\ 
78.4547 & 0.5233 & 0.0175 & 0.8705 & 0.0111 \\ 
56.0514 & 0.3189 & 0.0161 & 0.8977 & 0.0160 \\ 
24.0361 & 0.5432 & 0.0236 & 0.9464 & 0.0395 \\ 
42.5793 & 0.4410 & 0.0204 & 0.9376 & 0.0221 \\    
 \hline
\end{tabular}\caption{First five quantized points of the GMM estimators of the measure distortions parameters $(c,\gamma,b,a)$ for SPY.}
\label{GMMParamStats}
\end{table}

It is worth noting that, even more than in the case of calibration to option prices, the parameter $a$ for the DM estimators has no significance, which implies that $\Lambda_+$ is dominated by $\Lambda_-$ (see Figure \ref{MeasureDistortions}), as $\tfrac{b}{c}>a$, and that SPY's ask price is based on uncertainty on potential losses, while SPY's bid price on that of potential gains. With GMM estimators, instead, $a\approx b/c$ and the treatment of gains and losses is balanced under both distortions, in the sense that the maximum reached by $\Lambda_+$ and $\Lambda_-$ is the same. We observe that such balanced result obtained via GMM estimator is in line with typical assumptions in the literature on estimation of distortion parameters (as for instance in \cite{MadanEntropy} and \cite{MadanHDMT}). 

\begin{figure*}
		\centering
        \begin{subfigure}[b]{0.45\textwidth}
            \centering
            \includegraphics[width=\textwidth]
            {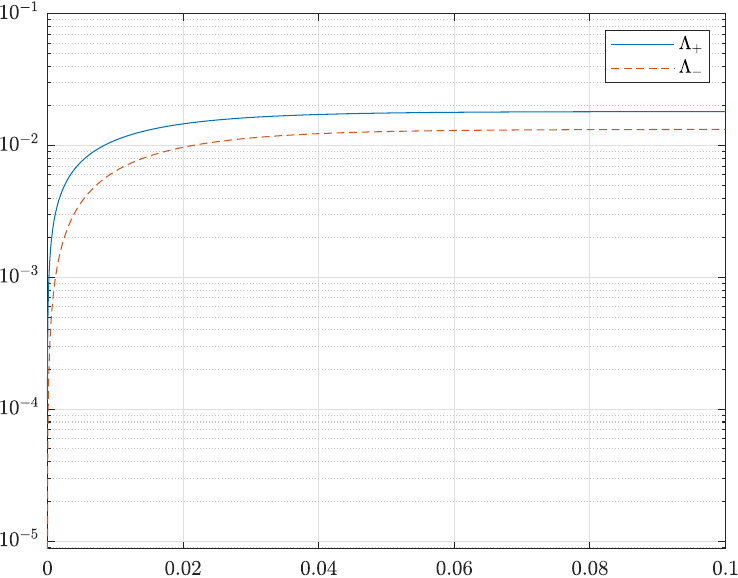}  
            \caption{{}}
        \end{subfigure}
        \begin{subfigure}[b]{0.45\textwidth}
            \centering
            \includegraphics[width=\textwidth]
            {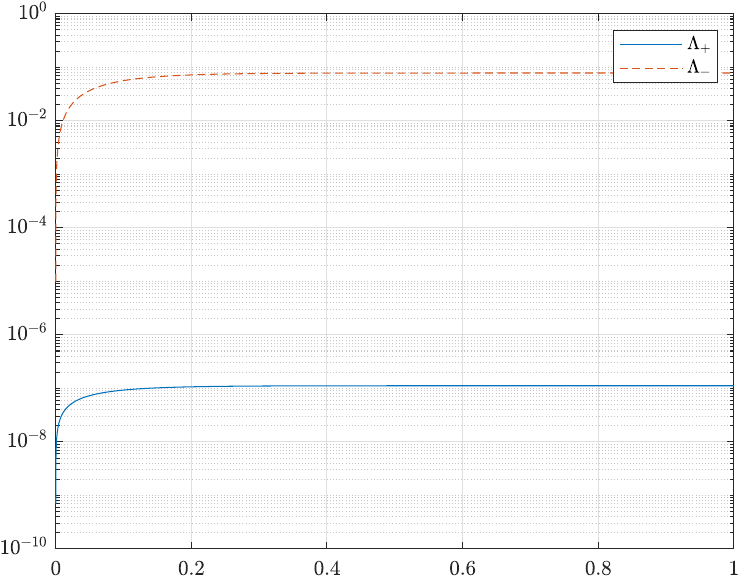}  
            \caption{{}}
        \end{subfigure}
		\caption{Measure distortions, in logplot, for SPY based on first quantized point of (A) GMM and (B) DM estimators of the measure distortion parameters.}\label{MeasureDistortions}
\end{figure*} 

\begin{remark} In a static setting, consistency between buying a claim $C$ and selling the claim $-C$ requires that the probability distortions that define the Choquet expectation of the upper valuation must be chosen so that the losses distortion is the dual of the gains distortion. In the continuous time limit, instead, the distortions $\Lambda_+$ and $\Lambda_-$ are no longer linked, as one can see by inspecting the proof of Theorem 5.2 in \cite{MadanPistoriusStadje}. This additional flexibility allows the unbalanced treatment of gains and losses.
\end{remark}

\item \textbf{Upper Valuations based on GMM Estimated Parameters is Smaller}

Table \ref{EmpiricalVerification} shows the annualized average of 5-day upper, mid and lower rates of return.

\begin{table}[h!]
\centering
\begin{tabular}{| c | c | c | c || c | c | c | c |} 
 \hline
 ETF & $\overline{\mu}$ & $\mu$ & $\underline{\mu}$ & ETF & $\overline{\mu}$ & $\mu$ & $\underline{\mu}$\\
 \hline
XLB & 4.16 & 4.36 & 4.41 & XLP & 6.98 & 7.05 & 7.08 \\  
XLE & -4.90 & -4.79 & -4.62 & XLU & 4.38 & 4.45 & 4.44 \\ 
XLF & 1.48 & 1.82 & 2.04 & XLV & 10.12 & 10.20 & 10.22 \\ 
XLI & 6.94 & 7.04 & 7.04 & XLY & 12.81 & 13.03 & 13.12 \\ 
XLK & 13.36 & 13.51 & 13.52 & SPY & 8.11 & 8.25 & 8.28 \\  
 \hline
\end{tabular}
\caption{Annualized averages (in percentage points) over the period 2010-2020 of upper, mid and lower logarithmic returns for 10 sector ETFs and SPY. }
\label{EmpiricalVerification}
\end{table}

\noindent \underline{Facts From Table \ref{EmpiricalVerification}}:

\noindent On average over the period considered,
\begin{itemize}
\item $\overline{\mu}<\mu<\underline{\mu}$, 
\item $\underline{\mu}-\mu<\mu-\overline{\mu}$
\end{itemize}
for each of the ETFs considered.

\noindent \underline{Consequences From Table \ref{EmpiricalVerification}}
\begin{itemize}
\item Table \ref{EmpiricalVerification} provides an empirical confirmation of inequality (\ref{DriftIneq});
\item by equations (\ref{PricingEquations}), and based on Table \ref{EmpiricalVerification}), one would expect that, for both DM and GMM estimated measure distortion parameters,
\begin{align}\label{DriftIneq2}
\int_{\R\backslash\{0\}}(e^y-1)\overline{\psi}_{\Lambda}(y)dy>
\int_{\R\backslash\{0\}}(e^y-1)\underline{\psi}_{\Lambda}(y)dy.
\end{align}
\end{itemize}

To check (\ref{DriftIneq2}) we computed upper, mid and lower drifts implied by the measure distortions parameters estimated via DM and GMM. See Table \ref{ExpectedGainLoss}.

\begin{table}[h!]
\centering
\begin{tabular}{| c | c | c | c | c | c | c |} 
 \hline
 & \multicolumn{3}{c|}{DM} & \multicolumn{3}{c|}{GMM}\\
 \hline
\makecell{\% of Points Represented} & $\overline{\mu}$ & $\mu$ &  $\underline{\mu}$ & 
 $\overline{\mu}$ & $\mu$ &  $\underline{\mu}$ \\
\hline
10.40 $\%$ &-7.16 & 7.31 & 8.04 & 3.02 & 7.31 & 8.03 \\ 
9.50 $\%$ &-0.23 & 7.31 & 7.68 & 2.12 & 7.31 & 7.65 \\ 
8.59 $\%$ &-8.74 & 4.68 & 5.62 & 1.63 & 4.68 & 5.61 \\ 
8.14 $\%$ &-4.47 & 6.17 & 7.24 & 3.05 & 6.17 & 7.23 \\ 
7.91 $\%$ &1.84 & 4.67 & 7.20 & 2.38 & 4.67 & 7.16 \\ 
\hline \hline
 \makecell{Weighted Average} & -4.31 & 4.92 & 6.15 & 2.19 & 4.92 & 6.16\\
\hline
\end{tabular}\caption{Upper, mid and lower drifts computed based on DM (left) and GMM estimators, at the first five of sixteen quantized points. Mid drifts were computed based on estimated BG parameters. The weighted average is computed based on the percentage of the population represented by each point.}

\label{ExpectedGainLoss}
\end{table}
%&-2.44 & 2.51 & 8.02 & 2.78 & 2.51 & 7.95 \\ 
%&-7.22 & 1.84 & 6.09 & 0.92 & 1.84 & 6.34 \\ 
%&-4.14 & 3.70 & 4.42 & 1.68 & 3.70 & 4.23 \\ 
%&-7.40 & 6.24 & 6.53 & 3.11 & 6.24 & 6.59 \\ 
%&2.75 & 6.77 & 8.11 & 1.50 & 6.77 & 8.24 \\ 
%&-8.10 & 4.86 & 4.04 & 2.55 & 4.86 & 4.13 \\ 
%&-1.76 & 4.42 & 5.77 & 0.67 & 4.42 & 5.75 \\ 
%&-3.95 & 5.69 & 5.75 & 1.73 & 5.69 & 5.99 \\ 
%&2.05 & 2.47 & 8.30 & 1.83 & 2.47 & 8.28 \\ 
%&-7.12 & 1.90 & 2.70 & 2.65 & 1.90 & 2.73 \\ 
%&-1.82 & 1.79 & 3.75 & 2.02 & 1.79 & 3.53 \\

\noindent \underline{Facts from Table \ref{ExpectedGainLoss}}:
\begin{itemize}
\item Both GMM and DM estimators are consistent with inequality (\ref{DriftIneq2});
\item The differences between mid and upper drifts and between lower and mid drifts are much larger than those for the daily returns averages shown in Table \ref{EmpiricalVerification};
\item The lower drift estimate is approximately the same for both GMM and DM;
\item The upper drift estimate is substantially lower for DM than it is for GMM.
\end{itemize}

\vspace{2mm}

\noindent \underline{Consequences from Table \ref{ExpectedGainLoss}}:
\begin{itemize}
\item DM and GMM estimators try to fit more than just the first moment, and with only four measure distortion parameters, thus the estimates are different than those in Table \ref{EmpiricalVerification};
\item The lower driver is similar across DM and GMM estimations, but the DM based upper driver is smaller than the GMM one; hence, DM implied upper valuations are higher than GMM implied ones.
\end{itemize}

\vspace{3mm}

\item \textbf{Low Correlation between GMM's Lower Driver and Lower Return.}

The correlation between the time series of upper drivers computed every 5 day based on GMM estimated parameters and the time series of 1-year average of 5-day upper returns, is significantly higher, on average, than the correlation between DM estimated upper drivers and the 1-year average of 5-day returns. The correlations between lower returns and lower drivers is instead more balanced. See Figure \ref{CorrTable}.
% and Figure \ref{RCvsReturnsSPY} shows the estimated drivers for the SPY from 2010 through 2020.

%\noindent \underline{Facts from Figure \ref{Corr}}:
%\begin{itemize}
%\item the correlation between realized upper returns and upper DM estimated drivers is similar to that of lower returns and lower DM estimated drivers;
%\item the correlation between realized lower returns and lower GMM estimated driver is consistently dominated by that of upper returns and upper GMM estimated driver.
%\end{itemize}

%Thus, Figure \ref{Corr} provides additional evidence for the inability of GMM estimated distortions to sufficiently differentiate between uncertainty in the upper price process and that for the lower price process.
%On the other hand, there is a clear higher correlation between the GMM estimated upper and lower drivers than there is between the DM estimated ones, which again reflects the unbalanced treatment of gains and losses for the DM estimators.
This higher correlation of GMM implied upper drivers and average upper returns is explained by the fact that GMM estimators are not designed to capture information in upper returns statistics of high orders. In other words, there is  significant amount of large observations of 5-day upper returns that are averaged out (and thus lost) in the computation upper returns' moments of order up to the sixth.
%\begin{figure*}
%        \centering
%        \begin{subfigure}[b]{0.4\textwidth}
%            \centering
%            \includegraphics[width=\textwidth]
%            {corrGMM10511-eps-converted-to.pdf}  
%            \caption{{}}
%        \end{subfigure}
%        \begin{subfigure}[b]{0.4\textwidth}
%            \centering
%            \includegraphics[width=\textwidth]
%            {corrDM10511-eps-converted-to.pdf}  
%            \caption{{}}
%        \end{subfigure}
%		\caption{Difference between correlation of upper drivers and upper returns and lower drivers and lower returns for GMM estimators (A) and DM estimators (B).}\label{Corr}
%\end{figure*} 

\begin{table}[h!]
\centering
\begin{tabular}{| c | c | c | c | c |} 
 \hline
 & \multicolumn{2}{c|}{DM} & \multicolumn{2}{c|}{GMM}\\
 \hline
\makecell{quantile} 
& \makecell{Upper \\ Correlation} 
& \makecell{Lower \\ Correlation} 
& \makecell{Upper \\ Correlation} 
& \makecell{Lower \\ Correlation} \\
\hline
0.00 & -0.93 & -0.92 & -0.72 & -0.79 \\ 
0.25 & -0.59 & -0.54 & 0.02 & -0.47 \\ 
0.50 & -0.29 & -0.10 & 0.21 & -0.25 \\ 
0.75 & 0.06 & 0.33 & 0.44 & -0.01 \\ 
1.00 & 0.68 & 0.85 & 0.76 & 0.60 \\ 
\hline
\end{tabular}\caption{Quantiles of the correlations between upper drift and average upper return and lower drift and average lower return for DM (left) and GMM (right) estimators.}
\label{CorrTable}
\end{table}

\vspace{3mm}

\item \textbf{Low Loss Tests Scenarios are consistent with Quantitative Easing}

How can we explain such high upper drift observations? By plotting the difference
\begin{align*}
\int_{\R\backslash\{0\}}(e^y-1)\overline{\psi}_{\Lambda}(y)dy-
\int_{\R\backslash\{0\}}(e^y-1)\underline{\psi}_{\Lambda}(y)dy
\end{align*}
of the two drivers, which are shown in Figure \ref{RCvsReturnsSPY}(A) and (B) for GMM and DM respectively, we can see that such distance almost triples between mid March and October 2020 when estimated with DM (this is seen better in Figure \ref{RCvsReturnsSPYII}(A)). Given the announcement on March 15 2020 by the Federal Reserve Board that it would ``Support the Flow of Credit to Households and Businesses'',\footnote{See \href{https://www.federalreserve.gov/newsevents/pressreleases/monetary20200315b.htm}{https://www.federalreserve.gov/newsevents/pressreleases/monetary20200315b.htm}.} one could then conjecture that it is such an announcement and its implementation that caused the increase in the upper driver with respect to the lower one. Because of the DM unbalanced treatment of the gain and loss processes, such increase corresponds to the market testing scenarios in which the weight given to the event that the exponential loss process be low is higher than that given to high exponential gain process realizations. More in general, Figure \ref{RCvsReturnsSPYII}(B) shows that a similar, albeit less pronounced, widening of the spread between DM estimated drivers also occurred in the proximity of each previous phase of quantitative easing.

\begin{figure*}
		\centering
        \begin{subfigure}[b]{0.45\textwidth}
            \centering
            \includegraphics[width=\textwidth]
            {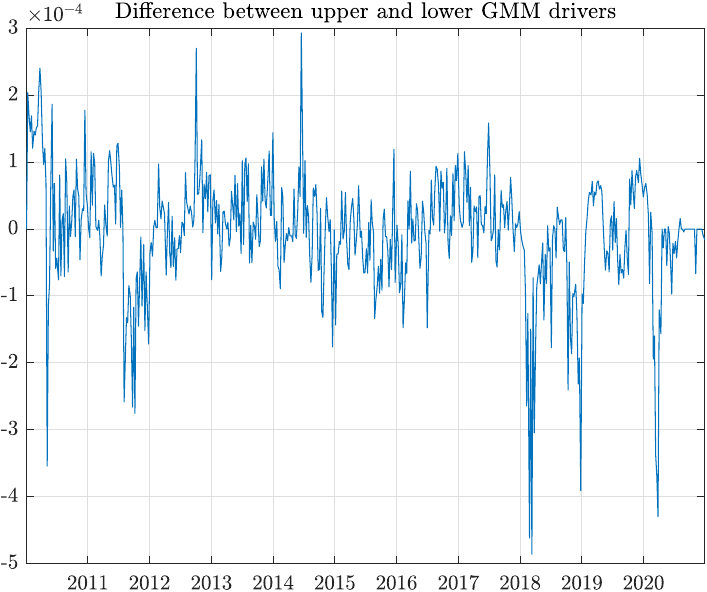}  
            %{SPY_20162020-eps-converted-to.pdf}
            \caption{{}}
        \end{subfigure}
        \begin{subfigure}[b]{0.45\textwidth}
            \centering
            \includegraphics[width=\textwidth]
            %{SPY_RCvsMktReturn_DigitalMoments1051-eps-converted-to.pdf}  
            {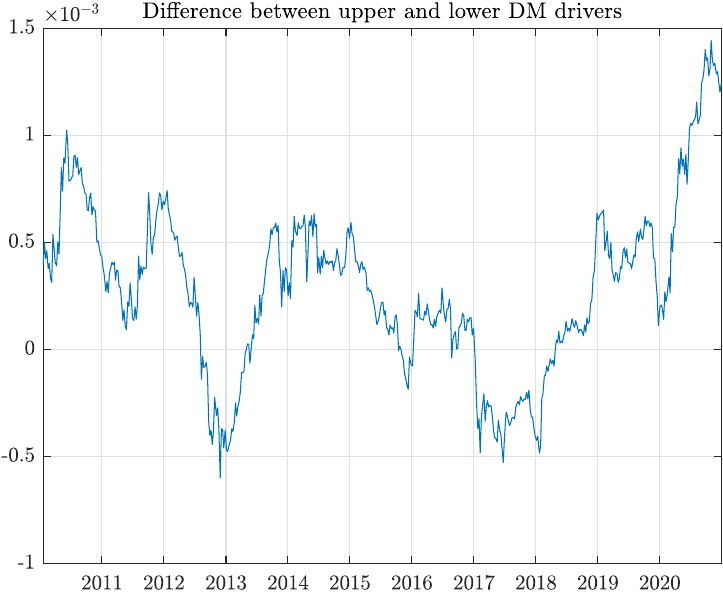}
            \caption{{}}
        \end{subfigure}
		\caption{Spread between drivers estimated with GMM (A) and DM (B).}\label{RCvsReturnsSPY}
\end{figure*}
\begin{figure*}
		\begin{subfigure}[b]{0.4\textwidth}
            \centering
            \includegraphics[width=\textwidth]
            %{SPY_RCvsMktReturn_DigitalMoments1051-eps-converted-to.pdf}  
            {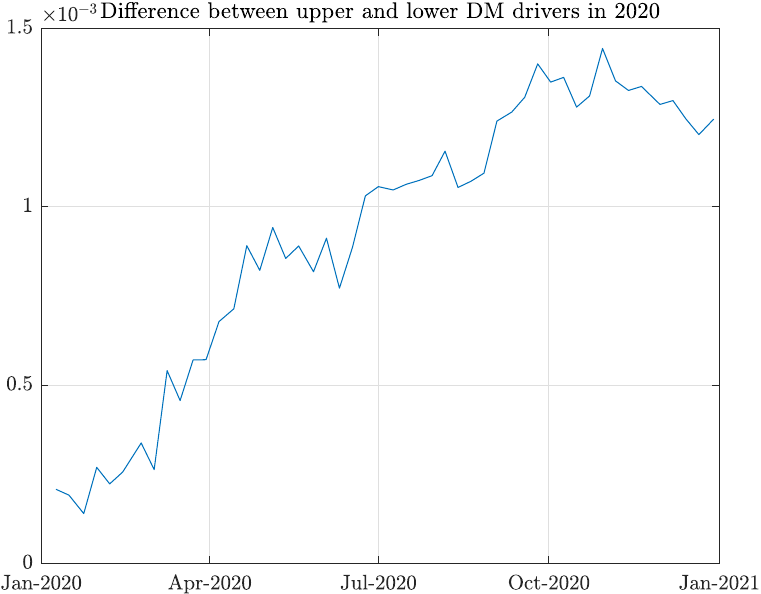}
            \caption{{}}
        \end{subfigure}       
%        \begin{subfigure}[b]{0.35\textwidth}
%            \centering
%            \includegraphics[width=\textwidth]
%            %{SPY_RCvsMktReturn_DigitalMoments1051-eps-converted-to.pdf}  
%            {SPY-eps-converted-to.pdf}
%            \caption{{}}
%        \end{subfigure}
        \begin{subfigure}[b]{0.4\textwidth}
            \centering
            \includegraphics[width=\textwidth]
            %{SPY_RCvsMktReturn_DigitalMoments1051-eps-converted-to.pdf}  
            {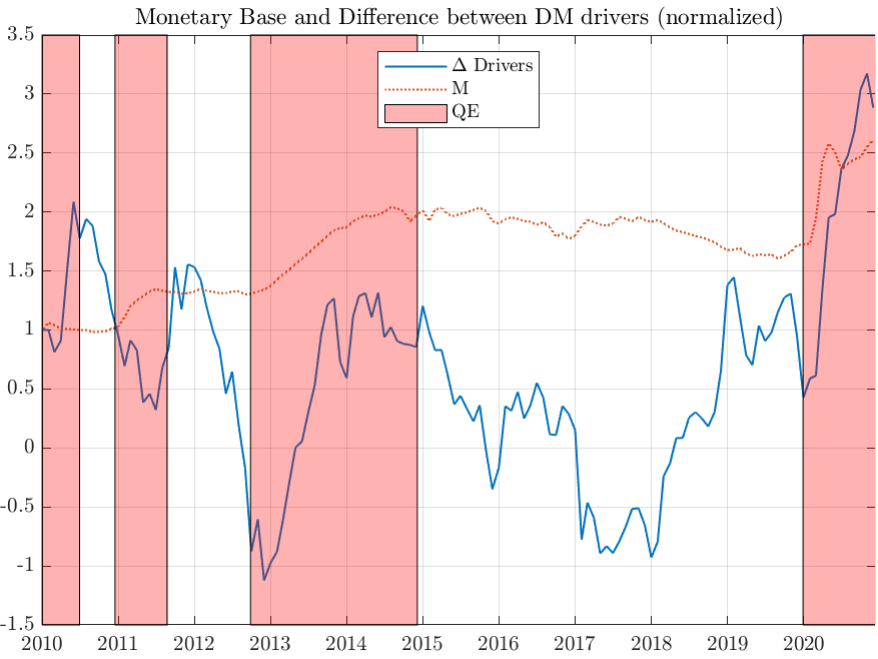}
            \caption{{}}
        \end{subfigure}  
		\caption{Figure (A): the DM spread during 2020. Figure (B): the DM spread and the total monetary base $M$ (source: FRED), with both series normalized to 1 on 1 January 2010. Shaded areas are the four phases of QE in the 2010-2020 decade.}\label{RCvsReturnsSPYII}
\end{figure*} 
On the other hand, the difference between GMM estimated drivers is, overall, an order of magnitude lower than that of DM estimated ones, as Figure \ref{RCvsReturnsSPY}(A) shows. This suggests that if only the first 6th moments of returns are matched, relevant tail events are averaged out and, thus, are not incorporated in the estimated drivers. 

%the augmented Dickey-Fueller test rejects the unit root model for the time series of the spread between GMM estimated drivers, shown in Figure \ref{RCvsReturnsSPY}(A). This suggests the nonstationarity of this time series, and, thus, that some information about future is picked up by the GMM drivers spread. The same test on the DM spread does not reject the null hypothesis of a unit root model.

\vspace{3mm}

\item \textbf{Higher Dispersion of Upper Valuations}

To visualize the difference between the two estimators, we computed the $L^1$ distance between the GMM and DM estimated densities for each day considered. The quantiles of the distances are summarized in Table \ref{L1DistanceQuantiles}. Figure \ref{AAStat} and \ref{AAStatDelta} show the GMM and DM  estimated densities and their difference as of 21 March 2020, a week after the above mentioned Federal Reserve Board's announcement. 

\begin{table}[h!]
\centering
\begin{tabular}{| c | c | c | c | c |} 
 \hline
 quantile & Upper Density & Lower Density \\
 \hline
0.25 & 21.2427 & 20.0399 \\ 
0.50 & 29.8376 & 45.0468 \\ 
0.75 & 38.6800 & 71.4766\\ 
 \hline
\end{tabular}\caption{Quantiles of the $L^1$ distance between the GMM vs DM estimated densities for the upper and lower distribution of returns.}
\label{L1DistanceQuantiles}
\end{table}

\begin{figure*}
\centering
        \begin{subfigure}[b]{0.45\textwidth}
            \centering
            \includegraphics[width=\textwidth]
            {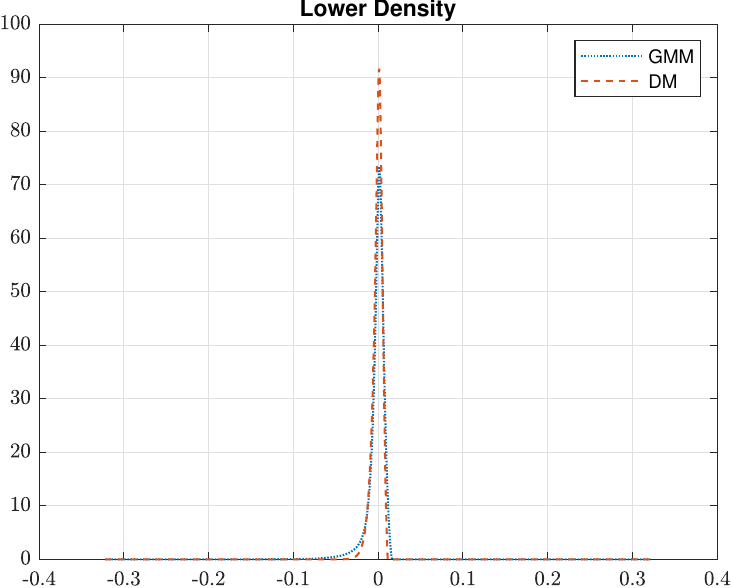}  
            \caption{{}}
        \end{subfigure}
        \begin{subfigure}[b]{0.45\textwidth}
            \centering
            \includegraphics[width=\textwidth]
            {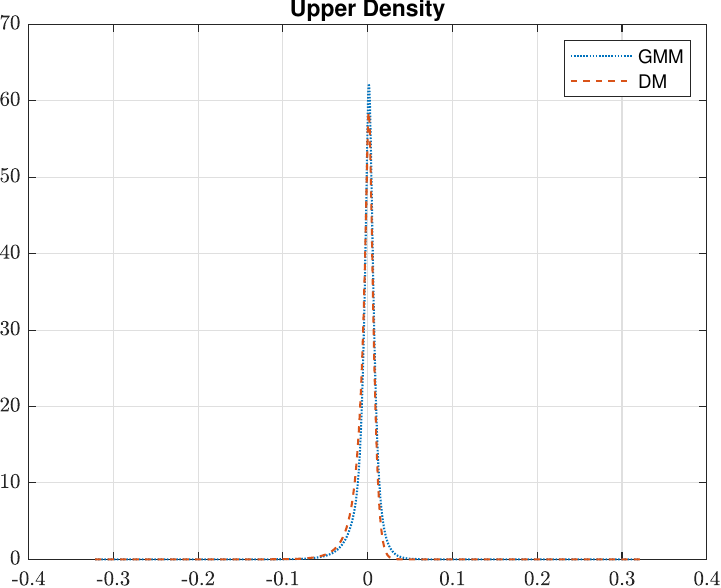}  
            \caption{{}}
        \end{subfigure}
		\caption{Upper and lower DM and GMM estimated densities as of 21 March 2020.}\label{AAStat}
\centering
        \begin{subfigure}[b]{0.45\textwidth}
            \centering
            \includegraphics[width=\textwidth]
            {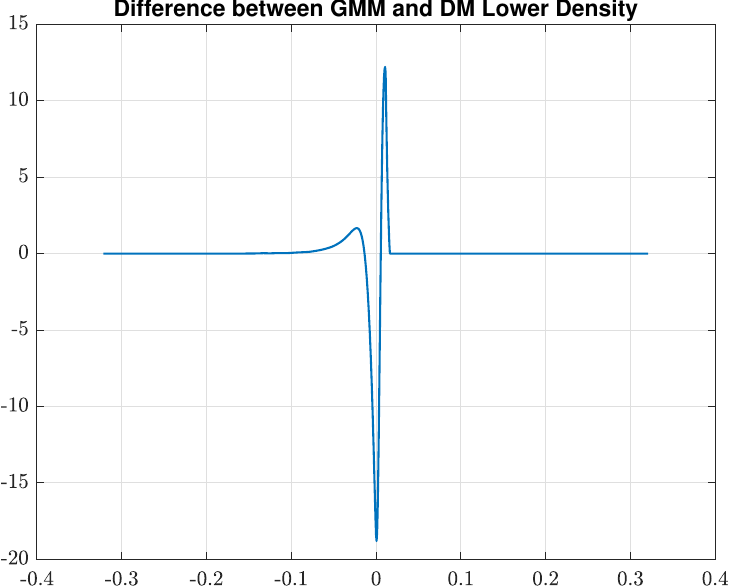}  
            \caption{{}}
        \end{subfigure}
        \begin{subfigure}[b]{0.45\textwidth}
            \centering
            \includegraphics[width=\textwidth]
            {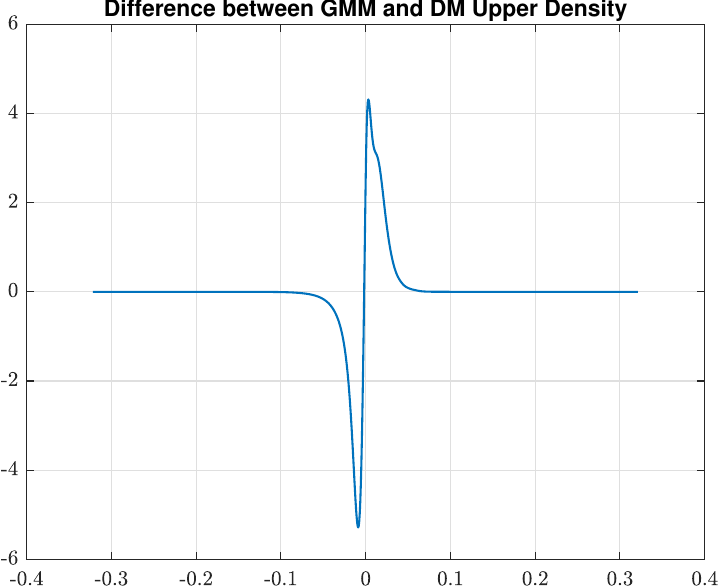}  
            \caption{{}}
        \end{subfigure}
		\caption{The distance between the estimated densities.}\label{AAStatDelta}		
\centering
        \begin{subfigure}[b]{0.45\textwidth}
            \centering
            \includegraphics[width=\textwidth]
            {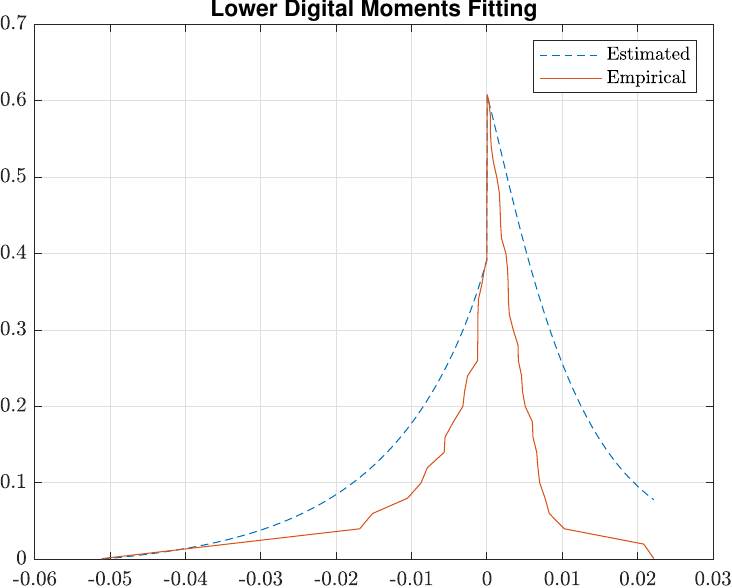}  
            \caption{{}}
        \end{subfigure}
        \begin{subfigure}[b]{0.45\textwidth}
            \centering
            \includegraphics[width=\textwidth]
            {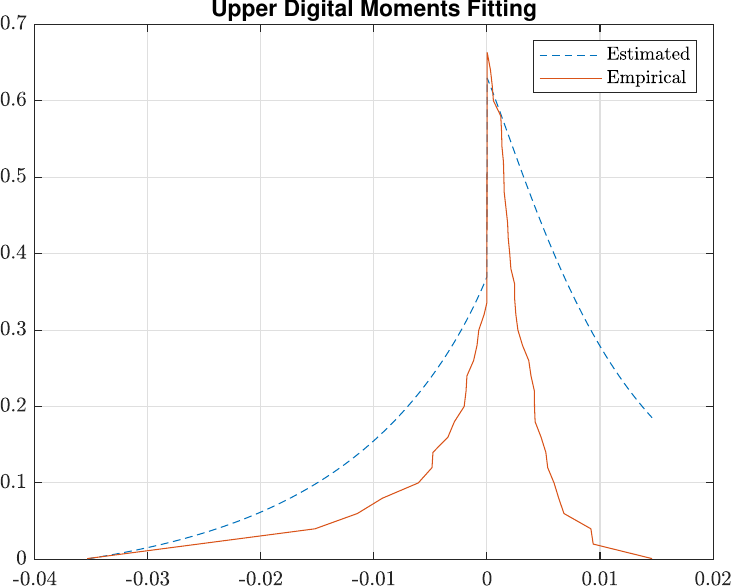}  
            \caption{{}}
        \end{subfigure}
		\caption{The fitting of the GMM and DM estimated upper (right) and lower (left) densities to the empirical survival functions $\overline{\Q}(\Lambda)(|X_T|>x)$ and $\underline{\Q}(\Lambda)(|X_T|>x)$.}\label{Fitting}		
\end{figure*}

The $L^1$ distance between the two densities on this date was actually within the interquantile range and close to the median, \footnote{Specifically, the $L^1$ distance between the two upper (resp. lower) distributions is 29.8 (resp. 67.6).} and their fitting to the empirical survival functions is similar (Figure \ref{Fitting}). However, one striking difference of the densities shown in Figure \ref{AAStat} is their level of dispersion, with, in particular, the DM estimated lower density being substantially less dispersed than the upper one compared to the GMM densities. In general, this feature holds true across all our daily estimates, as shown in Tables \ref{std}. 

\begin{table}[h!]
\centering
\begin{tabular}{| c | c | c | c | c | } 
 \hline
 & \multicolumn{2}{c|}{GMM} & \multicolumn{2}{c|}{DM}\\
 \hline
 quantile & \makecell{Upper \\ Density }& \makecell{Lower\\ Density } &
 \makecell{Upper \\ Density }& \makecell{Lower\\ Density } \\
 \hline
0.25 & 0.0121 & 0.0092 & 0.0126 & 0.0063 \\ 
0.50 & 0.0142 & 0.0153 & 0.0146 & 0.0078\\ 
0.75 & 0.0167 & 0.0240 & 0.0171 & 0.0098\\ 
 \hline
\end{tabular}\caption{Quantiles of the daily estimated standard deviation of $X_T$.}
\label{std}
\end{table}
\end{enumerate}

\section{Conclusions}

This paper provides a formula for the Radon-Nikodym derivative of a purely discontinuous L\'evy process $X$ under the extreme measures defined by continuous time Conic Finance. This result implies that continuous time Conic Finace valuations are not law invariant nor linear over comonotonic claims, as their static counterparts. Also, for one dimensional monotone claims the process $X$ is a L\'evy proces under the extreme measures, and its L\'evy density is explicit. This is useful in empirical studies as it allows the use in our nonlinear setting of estimation methodologies typically applicable only under the law of one price. In particular, we calibrated distortion parameters to forward looking option prices using the FFT method and for two different parametric families of the distortions, one of which is new and is seen as a generalization to a dynamic setting of the Wang transform construction. Furthermore, we estimate measure distortion parameters via GMM and DM based on historical equity prices. Both estimate capture market's higher uncertainty around upward potential of the SPY. However, such uncertainty appears significantly underestimated by the GMM. 

\section{Acknowledgment}
This paper is a revised version of a chapter of my doctoral dissertation. I am thankful to Prof. Madan, my supervisor, for the many suggestions and comments he provided while working on this paper. I also wish to thank two anonymous referees for providing suggestions that improved the paper, to Umberto Cherubini for a useful discussion, and to the participants to the April 2023 AMS Sectional Meeting in Stochastic Analysis and its Applications held at the University of Cincinnati.

\bibliographystyle{authordate1}
\bibliography{mybib_2011_pl}

\end{document}